\documentclass[a4paper,fleqn,usenatbib,twocolumn,doublespacing]{mnras}

\usepackage{newtxtext,newtxmath}
\usepackage{lipsum} % just for dummy text- not needed for a longtable

\usepackage[T1]{fontenc}
\usepackage{ae,aecompl}
\usepackage{graphicx}	% Including figure files
\usepackage{amsmath}	% Advanced maths commands
\usepackage{amssymb}	% Extra maths symbols
\usepackage{multicol}        % Multi-column entries in tables
\usepackage{bm}		% Bold maths symbols, including upright Greek
\usepackage{multirow}
\usepackage{enumitem}
\normalsize
\title[Bayesian Hierarchical Modelling IFMRs]{Bayesian Hierarchical Modelling of Initial-Final 
Mass Relations Across Star Clusters}

\author[Si, van Dyk, von Hippel et al.]{
Shijing Si,$^{1}$\thanks{Contact e-mail: \href{mailto:ss2913@ic.ac.uk}{ss2913@ic.ac.uk}}
David A. van Dyk,$^{1}$
Ted von Hippel,$^{2,3}$
Elliot Robinson,$^{4}$
\newauthor{Elizabeth Jeffery$^{5}$, and David C. Stenning$^{1}$}
\\
$^{1}$Statistics Section, Department of Mathematics, Imperial College London, London SW7 2AZ, UK\\
$^{2}$Physical Sciences Department, Embry-Riddle Aeronautical University, Daytona Beach, FL, USA\\
$^{3}$Institute of Astronomy, Madingley Road, Cambridge CB3 OHA, UK\\
$^{4}$Argiope Technical Solutions, FL, USA\\
$^{5}$Physics Department, California Polytechnic State University,  San Luis Obispo, CA, USA\\
\\
}
\date{Accepted XXX. Received YYY; in original form ZZZ}

\pubyear{2017}

\begin{document}
\label{firstpage}
\pagerange{\pageref{firstpage}--\pageref{lastpage}}
\maketitle

\begin{abstract}
The initial-final mass relation (IFMR) of white dwarfs (WDs) plays
an important role in stellar evolution.
To derive precise estimates of IFMRs and explore how they may vary among star clusters, 
we propose a Bayesian hierarchical model that pools
photometric data from multiple star clusters.
After performing a simulation study to show the benefits of the Bayesian 
hierarchical model, we apply this model to five star clusters: the Hyades, M67,
NGC 188, NGC 2168, and NGC 2477, leading to reasonable and consistent estimates of IFMRs
for these clusters.
We illustrate how a cluster-specific analysis of NGC 188 using its own photometric data can produce
an unreasonable IFMR since its WDs have a narrow range of zero-age main sequence
(ZAMS) masses.
However, the Bayesian hierarchical model corrects the cluster-specific analysis by borrowing strength
from other clusters, thus generating more reliable estimates of IFMR parameters.
The data analysis presents the benefits of Bayesian hierarchical modelling over conventional 
cluster-specific methods, which motivates us to elaborate the powerful statistical techniques
in this article.

\end{abstract}


%%%%%%%%%%%%%%%%%%%%%%%%%%%%%%%%%%%%%%%%%%%%%%%%%%

%%%%%%%%%%%%%%%%% BODY OF PAPER %%%%%%%%%%%%%%%%%%
\section{Introduction}
%\subsection{}

The initial-final mass relation (IFMR) provides a mapping
between the zero-age main sequence (ZAMS) mass of a star to its white dwarf (WD) mass and is
vital to an understanding of mass loss during stellar evolution.
Many researchers have investigated the IFMR using data from
different star clusters, leading to numerous
versions of the IFMR. 
For instance, \citet{williams2004}
presented an empirical IFMR based on spectroscopic analysis of seven massive WDs in 
NGC 2168 (M35).
\citet{kalirai2005the} presented $24$ new faint WDs in the open cluster NGC 2099 and
determined an IFMR based on their high turnoff mass ($\sim2.4 M_{\sun}$).
\citet{catalan2008the} explored the application of common proper motion pairs to
improve the IFMR.
 \citet{salaris2009} provided an empirical estimate
of the IFMR using published results of WD properties in ten clusters. 
\citet{williams2009} probed the empirical IFMR using
WDs in the open cluster NGC 2168 (M35) at the high-mass end of the relation.
\citet{cummings2016two} observed a sample of $10$ WD candidates in the open cluster NGC 2323 and
 investigated a linear IFMR for high-mass ($\geq0.9 M_{\sun}$) WDs. 
By contrast, \citet{jeffery2011the} and \citet{zhao2012the} studied the IFMR in the low ZAMS mass range of
$1-2 M_{\sun}$. 
\citet{andrews2015ifmr} identified $65$ new wide double WDs and used them to 
constrain the IFMR.

\citet{stein2013} treated the parametrised IFMR 
as cluster-specific parameters and developed simultaneous
 principled Bayesian estimates of all cluster-specific
parameters, including those describing the IFMR. 
In addition, \citet{stein2013} detected the disagreement of 
IFMRs from the Hyades, NGC 2168, and NGC 2477,
 which might be caused by many factors such as observation errors or
  metallicity differences among these clusters.
In this paper, we approach the possible variation of IFMRs for different clusters with
 a Bayesian hierarchical model, which
 on average produces more accurate estimates of the IFMR(s).
 
Bayesian hierarchical modelling \citep{gelman2006, gelman2013}
is a statistical method that simultaneously fits object-specific parameters for
multiple objects by pooling their data under one overall model.
 The resulting estimates from hierarchical models are
 shrinkage estimates \citep{si2017a} that
 generally have better statistical properties than
 do their unpooled counterparts.   
Bayesian hierarchical models have been used in numerous projects in astrophysics
\citep[e.g.,][]{jiao2016efficiency,shariff2016bahamas,kaisey2016the,si2017a, si2017sensitivity, si2018}. 
In the context of constraining the IFMR, \citet{andrews2015ifmr}
 pooled 142 wide double WDs in a hierarchical model.

In this paper we propose a Bayesian hierarchical model for cluster IFMRs, show how this model can be fit using 
existing software, and use a suite of simulation studies to verify the statistical advantages of the resulting IFRM 
estimates. 
We aim to perform a comprehensive analysis of the IFMR by combining multiple star clusters into
a hierarchical model. This allows us to simultaneously obtain better estimates of
each cluster's IFMR and to estimate the intrinsic variance of cluster-specific IFMRs.
We apply the Bayesian hierarchical model using data from five clusters:
the Hyades, M67, NGC 188, NGC 2168 and
NGC 2477. We obtain the shrinkage estimates of IFMR parameters for these five clusters. 

The paper is organised as follows. Section \ref{sec:stat} summarises the cluster-specific Bayesian model
for cluster parameters introduced by \citet{stein2013} and proposes a hierarchical model to
simultaneously fit multiple clusters. Section \ref{sec:sim} presents a simulation study and demonstrates the 
advantages
of the hierarchical model. In Section \ref{sec:data}, we analyse five clusters via both the
cluster-specific and hierarchical approaches, and illustrate the advantages of the
latter approach. Section \ref{sec:sen} covers the sensitivity analysis of the prior distribution
used in the hierarchical model and membership of WDs in the cluster M67.
The conclusion and discussion of the use of our
statistical technique appears in Section \ref{sec:con}.

\section{Statistical Models} \label{sec:stat}

In this Section, we review the Bayesian approach \citep{stein2013} to fit
cluster-specific IFMR parameters and
propose a hierarchical model that allows us to
combine data from multiple clusters to simultaneously
improve the estimate of the cluster-specific IFMR
parameters and to explore
the variability among IFMRs for different clusters.

\subsection{Cluster-specific Analyses}\label{sec:case}

\citet{stein2013} develop a Bayesian approach for cluster parameters such as age, metallicity, 
and distance modulus while simultaneously estimating the IFMR for that cluster.
They estimate the IFMR and other cluster parameters using a
state-of-the-art Markov chain Monte Carlo (MCMC) algorithm and
implement their methods using
the software package BASE-9 \citep{von2006inverting,degennaro2009inverting,van2009}.
BASE-9, short for Bayesian Analysis of Stellar Evolution with 9 parameters, deploys MCMC
techniques to perform reliable Bayesian analysis for
physical properties including age, distance modulus, metallicity and mass,
based on the photometry of stars in a star cluster. 
\citet{stein2013} fit one cluster at a time, i.e., {\textit cluster-specific analysis},
so that
each cluster has its own fitted IFMR.

In this paper we adopt a similar mathematical notation to that of \citet{stein2013},
 while the subscript is extended to accommodate
multiple star clusters. 
Suppose we have photometry for $K$ star clusters, along with measurement errors.
The number of stars in each cluster can vary, as can the number of photometric magnitudes
observed for each cluster or even for the stars within the clusters. 
We use $k$ to index clusters and $i$ to index stars within clusters.
Without loss of generality, we assume the number of stars within each cluster is $N$ and
 that the observed photometry vector for
 star $i$ within cluster $k$ is $\bm{X}_{ki}$,
with known measurement variance-covariance matrix $\bm{\Sigma}_{ki}$.
We assume that age ($\theta_{\rm age}$), metallicity ($\theta_{\rm [Fe/H]}$),
distance modulus ($\theta_{\rm m-M_{V}}$), and
absorption ($\theta_{\rm Av}$) are common to all stars in each cluster, and we
denote them together as
  $\bm{\Theta}_{k}=(\theta_{{\rm age},k}, \theta_{{\rm[Fe/H]},k}, 
  \theta_{{\rm m-M_{V}},k}, \theta_{{\rm Av},k})$.  We denote the parameters
  describing the IFMR of cluster $k$ by $\bm{\alpha}_{k}$; below  we use a linear
  IFMR model so each $\bm{\alpha}_{k}$ consists of a intercept and a slope. 
  Since $\bm{\alpha}_{k}$ is the same for all WDs in cluster $k$,
  we treat $\bm{\alpha}_{k}$ as a cluster parameter.
   We denote the ZAMS mass of star $i$ within cluster $k$ as $M_{ki}$.
  Also, any star in the dataset may be a field star, i.e., not a member of a specific cluster. 
We define $\bm{Z}_{k}=(Z_{k1}, \ldots, Z_{kN})$,
where $Z_{ki}=1$ if
 star $i$ observed on the field of the
 sky with cluster $k$ is indeed a cluster member, otherwise $Z_{ki}=0$.
 (Of course $Z_{ki}$ is unobserved and must be estimated.)
  See Table \ref{tab:params} for
  a summary of the model parameters.
   
  \begin{table}
 \caption{Cluster and stellar parameters.}
 \label{tab:params}
 \begin{tabular}{ll}
  \hline\hline
  Cluster parameters & \\
  \hline
  $\theta_{{\rm age}, k}$ & $\log_{10}$ age of the cluster $k$ \\ 
 $\theta_{{\rm [Fe/H]}, k}$ &  metallicity of the cluster $k$ \\ 
 $\theta_{{\rm m-M_{V}}, k}$ & distance modulus of the cluster $k$ \\ 
 $\theta_{{\rm Av}, k}$ & absorption in $V$-band mag.\\
 &   of the cluster $k$ \\ 
 $\bm{\alpha}_{k}$ & IFMR parameters of the cluster $k$ \\\hline
   Stellar parameters &  \\\hline
   $M_{ki}$ &  the initial mass of the observed \\
                  & star $i$ of the cluster $k$\\
   $Z_{ki}$ & indicator for the membership of the \\
     & observed star $i$ in the cluster $k$\\
   \hline
\end{tabular}
\end{table}
 
We parametrise the IFMR of cluster $k$ as a linear form
 \begin{equation}\label{eq:linrIFMR} 
M_{{\rm WD},ki}=\alpha_{k0} + \alpha_{k1}(M_{ki}- 3.0)~\mbox{for WD $i$ in cluster $k$},
\end{equation}
where $\bm{\alpha}_{k}=(\alpha_{k0}, \alpha_{k1})$ are the intercept and slope parameters,
and $M_{WD, ki}$ is
the mass of WD $i$ within cluster $k$. 
Specifically $\alpha_{k0}$ is the WD mass of a star in cluster $k$ with
progenitor ZAMS mass equal to
$3.0~M_{\sun}$. For every additional increment of $1.0~M_{\sun}$ in ZAMS
mass, we expect the WD mass to increase by $\alpha_{k1}$. 

Though we have distinct evolution models for MS/RG and WD stars,
we denote them indistinguishably by $\bm{G}(\cdot)$, which comprises MS/RG
evolution models, WD cooling models, WD atmosphere models, and IFMR models.
Because the expected photometric magnitudes of WDs depend on the WD masses,
$\bm{G}(\cdot)$ must incorporate $\bm{\alpha}_{k}$. Thus, 
 for the reminder of this article, the stellar evolution model $\bm{G}(\cdot)$,
is viewed as a function of $\bm{\alpha}_{k}$
in addition to $\bm{\Theta}_{k}$ and $M_{ki}$.
Due to the computational complexity of stellar evolution models, in practice we employ
a computer-based model to evaluate $\bm{G}(\cdot)$.

The cluster-specific model for cluster $k$ is
\begin{equation}\label{eq:case1}
   \bm{X}_{ki}\mid(\bm{\Theta}_{k}, M_{ki}, \bm{\alpha}_{k}) \sim{\rm N}(\bm{G}(\bm{\Theta}_{k},
  M_{ki}, \bm{\alpha}_{k}), \bm{\Sigma}_{ki}),~~{\rm if}~Z_{ki}=1,
  \end{equation}
 where ${\rm N}$ is a multivariate Gaussian distribution with mean $\bm{G}(\cdot)$ and 
 variance-covariance matrix $\bm{\Sigma_{ki}}$, and
 $\bm{G}(\bm{\Theta}_{k}, M_{ki}, \bm{\alpha}_{k})$ is the predicted vector of 
 photometric magnitudes
for star $i$ within cluster $k$.
Eq. \ref{eq:case1} summarises the probabilistic relationship between the
photometry of stars that are members of cluster $k$ (i.e., stars with $Z_{ki}=1$) and the model parameters. 
If a star in dataset $k$ is a field star (i.e., $Z_{ki}=0$),
 we assume that its magnitudes are uniformly distributed on a hyper-rectangle which includes
 the full range of observed magnitudes of stars for that field.
 We use $p_{\rm field}(\cdot)$ to denote the distribution of photometric magnitudes for field stars,
 which is simply the reciprocal of the volume of the hyper-rectangle.
 To be specific,
 for example, stellar cluster $k$ has photometric magnitudes in the $U, B, V$ filters. Then 
 we find the range of $U, B, V$ by using the maximum value of each filter minus
its minimum and denote them $\ell_{U}, \ell_{B}$ and $\ell_{V}$.
If a star with magnitudes $\bm{X}_{ki}$ is a field star, its likelihood is
 $p_{\rm field}(\bm{X}_{ki})=\frac{1}{\ell_{U}\ell_{B}\ell_{V}}$.
 Though uniform model for field stars is unrealistic, 
 \citet{stenning2016bayesian} used a simulation study to demonstrate that the simple 
 but physically unrealistic model can nevertheless identify field stars with a high level of accuracy.
  For details, refer to
 Page 10 of \citet{stenning2016bayesian}.
 Therefore, in this research I use uniform model for field stars in each cluster.
 
Taken together, this means that the likelihood function for cluster $k$ is
\begin{align}
&{L}(\bm{M}_{k}, \bm{\Theta}_{k}, \bm{Z}_{k}, \bm{\alpha}_{k}\mid\bm{X}_{k}, \bm{\Sigma}_{k})=
\prod_{i=1}^{N}\bigg[\frac{Z_{ki}}{\sqrt{|2\pi\bm{\Sigma}_{ki}|}}\times\\ \nonumber
&\exp\bigg(-\frac{1}{2}\Big(\bm{X}_{ki}-\bm{G}(M_{ki}, \bm{\Theta}_{k}, \bm{\alpha}_{k})\Big)^{\top}\bm{\Sigma}_{ki}^{-1}\Big(\bm{X}_{ki}-\bm{G}(M_{ki}, \bm{\Theta}_{k}, \bm{\alpha}_{k}\Big)\bigg)\\\nonumber
&+ (1-Z_{i}){p}_{\rm field}(\bm{X}_{ki})\bigg],\label{eq:like}
\end{align}
where $\bm{M}_{k}=(M_{k1}, \ldots, M_{kN}),
\bm{X}_{k}=(\bm{X}_{k1},\ldots,\bm{X}_{kN})$, and
$\bm{\Sigma}_{k}=(\bm{\Sigma}_{k1}, \ldots, \bm{\Sigma}_{kN})$.
The prior distribution for the parameters is assumed to be
\begin{align}
&{p}(\bm{\Theta}_{k}, \bm{\alpha}_{k}, \bm{M}_{k}, \bm{Z}_{k})=
{p}(\bm{\Theta}_{k}){p}(\bm{\alpha}_{k}){p}(\bm{M}_{k})p(\bm{Z}_{k})\\ \nonumber
=&{p}(\theta_{{\rm age},k}){p}(\theta_{{\rm [Fe/H]},k}){p}(\theta_{m-M_{V},k}){p}(\theta_{{\rm Av},k})
{p}(\bm{\alpha}_{k})\times\\\nonumber
&~~\prod_{i=1}^{N}{p}(M_{ki})p(Z_{ki}).\label{eq:prior}
\end{align}
Specifically, for 
$\theta_{{\rm age},k}, \theta_{{\rm[Fe/H]},k}, \theta_{{\rm m-M_{V}},k}~\mbox{and}~\theta_{{\rm Av},k}$ we use
independent Gaussian prior distributions, with means set in accordance with
 recently published fits and variances chosen to be
reasonably non-informative. 
 In so doing, we eliminate the influence of prior distributions
in our analyses.
For the IFMR intercept $\alpha_{k0}$ we use a uniform prior distribution on the real line. 
 For the IFMR slope $\alpha_{k1}$ we use a uniform distribution on the positive part of the real line, which
excludes the possibility of a decreasing IFMR. The prior probability of star $i$ being
a member of cluster $k$, $p(Z_{ki}=1)$ is set based on the best available external information, typically
using proper motions and/or radial velocities.
Finally, we
use one version of the initial mass
 function (IMF) of \citet{miller1979} as the prior distribution of the ZAMS mass for star $i$, i.e.,
$$p(\log_{10}(M_i))\propto\exp\bigg(-\frac{1}{2}\Big(\frac{\log_{10}(M_i) + 1.02}{0.677}\Big)^2\bigg),$$
truncated to 0.1 $M_{\odot}$ to 8 $M_{\odot}$.
The lower truncation is due to the fact that an initial mass of less than about 0.1 $M_{\odot}$ is 
not sufficient to initiate the fusing of hydrogen into helium necessary to form a star. 
The upper truncation is because the star clusters we study are sufficiently 
old that any stars with an initial mass above 8 $M_{\odot}$ would have used up 
their nuclear fuel long ago and become a neutron star or black hole, 
and thus would not be included in our observed data \citep{van2009}.

In their cluster-specific analysis of cluster $k$, \citet{stein2013} based statistical
inference, including parameters' estimates and error bars, on 
the joint posterior distribution,
\begin{equation}\label{eq:case.post}
\begin{split}
&{p}(\bm{\Theta}_{k}, \bm{M}_{k}, \bm{Z}_{k}, \bm{\alpha}_{k}|\bm{X}_{k}, \bm{\Sigma}_{k})\\
&\propto{p}(\bm{\Theta}_{k}, \bm{\alpha}_{k}, \bm{M}_{k}, \bm{Z}_{k})
{L}(\bm{M}_{k}, \bm{\Theta}_{k}, \bm{Z}_{k}, \bm{\alpha}_{k}| \bm{X}_{k}, \bm{\Sigma}_{k}).
\end{split}
\end{equation}
BASE-9 can draw a reliable sample for all parameters from their joint posterior distribution in
Eq. \ref{eq:case.post}.
This is a cluster-specific study of the IFMRs because the fits of the
IFMR parameters only rely on data from one cluster.
We aim to perform a comprehensive analysis of the IFMR by combining multiple star clusters into
a hierarchical model. This allows us to simultaneously obtain better estimates of
each cluster's IFMR and to estimate the intrinsic variance of cluster-specific IFMRs. 

\subsection{Hierarchical Model}\label{sec:hier1}

In this section, we describe how to pool data from multiple star clusters using a hierarchical model and
how we fit this comprehensive model.
For $K$ star clusters, our hierarchical model is
\begin{equation}\label{eq:hier}
\left \{\begin{aligned}
   \bm{X}_{ki}\mid(\bm{\Theta}_{k}, M_{ki},\bm{\alpha}_{k})&\sim{\rm N}\Big(\bm{G}(\bm{\Theta}_{k},
  M_{ki}, \bm{\alpha}_{k}), \bm{\Sigma}_{ki}\Big), ~~\text{if}~~Z_{ki}=1,\\
 {\rm where}~~\bm{\alpha}_{k}&\sim{\rm N}(\bm{\gamma},~~\bm{\Gamma}), \, k= 1, \ldots, K.
 \end{aligned} \right.
\end{equation}
For field stars ($Z_{ki}=0$), $\bm{X}_{ki}$ is uniformly distributed on the aforementioned hyper-rectangle.
We set prior distributions on $\bm{\Theta}_{k}, \bm{M}_{k}, \bm{Z}_{k}, k=1, \ldots, K$
as in the aforementioned cluster-specific analysis. We assume that 
IFMRs of different clusters follow a common bivariate normal distribution,
which corresponds to the expectation
that the IFMRs of different clusters, although not identical, are similar.
The only new parameters in the hierarchical model in Eq. \ref{eq:hier}
are $\bm{\gamma}$, the mean of the IFMR intercept and slope, and 
the $\bm{\Gamma}$ which is the variance-covariance matrix of IFMR parameters among the clusters.
This assumption means that IFMR parameters of different clusters
are from the same bivariate normal population with mean $\bm{\gamma}$ and
variance-covariance matrix $\bm{\Gamma}$.
We must set prior distributions for $\bm{\gamma}$ and $\bm{\Gamma}$ and so we set
${p}(\bm{\gamma}, \bm{\Gamma})={p}(\bm{\gamma}|\bm{\Gamma}){p}(\bm{\Gamma})$ 
with $p(\bm{\gamma}\mid\bm{\Gamma})$ uniform on its range. For $\bm{\Gamma}$, we set
$$
\bm{\Gamma}\mid\lambda_{1}, \lambda_{2}\sim\mbox{Inverse-Wishart}
\Bigg(2\nu
\begin{pmatrix}
1/\lambda_{1} & 0\\
0 & 1/\lambda_{2}
\end{pmatrix},~~\nu+1\Bigg),
$$
where the {Inverse-Wishart}\footnote{
The Inverse-Wishart distribution is the conjugate prior distribution
 for the variance-covariance matrix of a multivariate normal
distribution. The Inverse-Wishart distribution is parametrised in terms of its
scale matrix $\bm{\Psi}$ and its degrees of freedom $\nu$;
its probability density function is 
$${p}(\bm{X} | \bm{\Psi}, \nu)=\frac{|\bm{\Psi}|^{\frac{\nu}{2}}}{2^{\frac{p\nu}{2}}
\Gamma_{p}(\frac{\nu}{2})}|\bm{X}|^{-\frac{\nu+p+1}{2}}e^{-\frac{1}{2}tr(\bm{\Psi}\bm{X}^{-1})},$$
where $\Gamma_{p}$ is the multivariate gamma function and $tr$ is the trace function.} 
 is the prior distribution for the variance matrix $\bm{\Gamma}$ given $\lambda_{1}$ and $\lambda_{2}$,
with $\lambda_{1}, \lambda_{2}\sim\mbox{Inverse-Gamma}(1/2,  1/5000)$\footnote{The Inverse-Gamma
is the reciprocal of of Gamma distribution, parametrised by its shape $\alpha$ and its rate $\beta$;
its density function is 
$$p(x | \alpha, \beta)=\frac{\beta^{\alpha}}{\Gamma(\alpha)}x^{-\alpha-1}\exp\Big(-\frac{\beta}{x}\Big),$$
where $\Gamma$ is the Gamma function.
}. It is sensible to take a diagonal scale matrix in the prior distribution of $\bm{\Gamma}$
because we parametrise the linear IFMR in Eq. \ref{eq:linrIFMR} in terms of 
$(M_{ki} - 3.0)$, where $3.0$ is near the average of the ZAMS masses of the
WDs in our clusters. This way of parametrisation decreases the correlation
between IFMR intercept and slope, simplifying computation of the hierarchical model. 
\citet{huang2013simple} suggests setting $\nu=2$ for a weakly informative prior
distribution on $\bm{\Gamma}$. In this paper, we set $\nu=2$ and
take a weakly informative distribution for $\bm{\Gamma}$, which
reduces the effect of the prior distribution and produces
estimates that mostly depend on the photometric data. 

We fit the hierarchical model Eq. \ref{eq:hier} in a Bayesian manner, and
it infers all parameters via their marginal posterior 
distributions by integrating out other parameters from their joint posterior distribution.
MCMC techniques are employed to simulate samples of all parameters. 
For details about the statistical inference of hierarchical models, see \citet[][]{gelman2013,si2017a}.
\label{sen:fb.def}
The joint posterior density for all parameters in Eq. \ref{eq:hier} is
\begin{equation}\label{eq:jp}
\begin{split}
&{p}(\bm{\gamma},\bm{\Gamma},\lambda_{1},\lambda_{2},\bm{\Theta}_{1}, \bm{M}_{1}, \bm{Z}_{1}, \bm{\alpha}_{1},\cdots, \bm{\Theta}_{K}, \bm{M}_{K}, \bm{Z}_{K}, \bm{\alpha}_{K}\\
&~~\mid\bm{X}_{1}, \bm{\Sigma}_{1}, \cdots, \bm{X}_{K}, \bm{\Sigma}_{K})\\
&={p}(\lambda_{1}){p}(\lambda_{2}){p}(\bm{\gamma},\bm{\Gamma}\mid\lambda_{1}, \lambda_{2})\times
\prod_{k=1}^{K}\\
&\bigg({p}(\bm{\Theta}_{k}){p}(\bm{M}_{k}, \bm{Z}_{k}){p}(\bm{\alpha}_{k}\mid\bm{\gamma},\bm{\Gamma}){L}(\bm{M}_{k}, \bm{\Theta}_{k}, \bm{Z}_{k}, \bm{\alpha}_{k}\mid\bm{X}_{k}, \bm{\Sigma}_{k})\bigg)
\end{split}
\end{equation}
In Appendix \ref{app:alg} we present a two-stage algorithm that draws a
reliable sample for parameters in the joint posterior in Eq. \ref{eq:jp}.
In the first stage, it draws a sufficient sample of
parameters from the cluster-specific analysis in Eq. \ref{eq:case.post}
to be used as the proposal distribution in a Metropolis-Hastings sampler
with target distribution equal to
 the hierarchical posterior distribution in Eq. \ref{eq:jp}. 
This strategy tackles the high-dimensional sampling problem in Eq. \ref{eq:jp} by taking 
advantage of the cluster-specific analyses.

\section{Simulation Study}\label{sec:sim}

To illustrate the statistical advantages of our hierarchical model,
we simulate $K=10$ star clusters with BASE-9 and we
recover their IFMRs via both cluster-specific and hierarchical analyses.
We simulate the cluster parameters using the distributions in Table \ref{tab:dists}.
These parameter values in Table \ref{tab:dists} are set to be similar to those of the observed
star clusters that we analyse in Section \ref{sec:data}. To mimic the errors of the observed photometry,
we compute the average standard deviations for filters $B, V$, and $I$ of the WDs in the
datasets analysed in Section \ref{sec:data} and use them as the corresponding errors in the simulated datasets. 
Specifically, the observed errors for $B, V, I$ are $\sigma_{B}=0.026, \sigma_{V}=0.035$, and
$\sigma_{I}=0.185$.

 \begin{table}
 \centering
 \caption{Parameter distributions for Simulated Clusters}
 \vspace{.2cm}
 \label{tab:dists}
 \begin{tabular}{l}
  \hline\hline
  $\theta_{{\rm age}, k}\sim{N}(9.0, 0.1^{2})$, \\ 
 $\theta_{{\rm [Fe/H]}, k}\sim{N}(-0.08, 0.05^{2})$, \\ 
 $\theta_{{\rm m-M_{V}}, k}=10.0$,  \\ 
 $\theta_{{\rm Av}, k}\sim{N}(0.51, 0.43^{2})$ truncated to positive values, \\ 
$\bm{\alpha}_{k}\sim{N}_{2}(\bm{\gamma}_{0}, \bm{\Gamma}_{0})$, with
\\
$
 \bm{\gamma}_{0}=\begin{pmatrix}
    0.722  \\
    0.107
  \end{pmatrix}$,
$
 \bm{\Gamma}_{0}=\begin{pmatrix}
    0.05^{2} & -0.2\cdot0.05\cdot0.02  \\
    -0.2\cdot0.05\cdot0.02 & 0.02^{2} 
  \end{pmatrix}$\\
\hline
\end{tabular}
\end{table}

After simulating the parameters $\theta_{{\rm age}, k}$,
 $\theta_{{\rm [Fe/H]}, k}$, $\theta_{{\rm m-M_{V}}, k}$, and $\theta_{{\rm Av}, k}$ of
 cluster $k$, we simulate photometric data for each star in the cluster
using BASE-9 \citep[see the BASE-9 User's Manual,][]{von2014bayesian}. 
For each cluster, we simulate $200$ MS/RG stars
brighter than $V=15$ and
 $10$ WDs.
  Subsequently, we recover the parameters of each cluster
 by fitting the simulated datasets with
 BASE-9 using the cluster-specific analysis 
 described in Section \ref{sec:case}. In so doing, we
 obtain a sample of the parameters for each cluster from its
 posterior distribution, see Eq. \ref{eq:case.post}.
For this paper, we
employ the \citet[][as updated at \url{http://stellar.dartmouth.edu/~models/}]{dotter2008} MS/RG models, 
\citet{montgomery1999evolutionary} WD interior models, \citet[][with updates posted online]{bergeron1995}
WD atmospheres, and the IFMR model in Eq. \ref{eq:linrIFMR}.

We repeat the data generation and parameter estimation 25 times,
record results from both the hierarchical and case-by-case methods.
We compare estimates in terms of two criteria: i.) root mean squared error (RMSE)
of point estimates, and 2.) actual versus nominal coverage probabilities of interval estimates.
When we require point estimates, we
utilise posterior means from MCMC
samples, and for interval estimates
 we use the 68.3\% credible intervals from posterior distributions
 by finding the 15.8\% and 84.1\% quantiles from the MCMC samples.

\begin{table*}
\centering
\caption{RMSE and coverage probabilities (CP) of 68.3\% credible intervals (CI) for IFMR parameters
from Bayesian hierarchical modelling and case-by-case analyses.}
  \vspace{.2cm}
\label{tab:ifmr.sum1}
\begin{tabular}{lllllllll}
\hline\hline
\multirow{2}{*}{Items}         && \multicolumn{3}{c}{Hierarchical} &  & 
\multicolumn{3}{c}{Case-by-case} \\ \cline{3-5} \cline{7-9} 
                               & & RMSE & CP &68.3\% CI &   & RMSE & CP &68.3\% CI \\ \hline
IFMR Const.  && 0.067 & 0.764 & $(0.734,0.792)$ && 0.202 & 0.8 & $(0.771,0.826)$ \\
IFMR Slope && 0.019 & 0.628 & $(0.595,0.660)$ && 0.331 & 0.844 & $(0.818,0.867)$ \\    \hline
\end{tabular}
\end{table*}

Table \ref{tab:ifmr.sum1} presents the RMSE of point estimates and the actual
 coverage probabilities of 68.3\% credible
interval estimates for IFMR constants and slopes using two methods: Bayesian hierarchical modelling
 and case-by-case.
The
68.3\% confidence intervals of the coverage probabilities are computed with 
the Clopper-Pearson exact method \citep{clopper1934use}.
From this table, the RMSE of hierarchical estimates of IFMR constants is 0.067, about a third of
that from the case-by-case method (0.202).
The performance of hierarchical modelling is even better on IFMR slope with its RMSE, 0.019, about $1/17$
of that from the case-by-case analysis.
In terms of interval estimates of IFMR parameters, the case-by-case (cluster-specific)
method has actual coverage probabilities, 80\% and 84.4\% for IFMR constant and slope respectively, 
higher than the nominal value, 68.3\%. 
The actual coverage probabilities of interval estimates from Bayesian hierarchical modelling, 76.4\% and 62.8\% 
for IFMR constant and slope respectively,
 are closer to the nominal value.
In summary, the estimates of IFMR parameters from
the hierarchical fits outperform that from case-by-case fits in terms of RMSE and coverage property.

 Here are population-level parameters:
the population mean of IFMR parameters, $\bm{\gamma}=(\gamma_{1}, \gamma_{2})$,
in which $\gamma_{1}$ and $\gamma_{2}$ are the mean of IFMR constants and slopes, respectively;
the population variance-covariance matrix of IFMR parameters, 
$$\bm{\Gamma} =
\begin{pmatrix}
\sigma_{1}^2 & \rho\sigma_{1}\sigma_{2} \\
\rho\sigma_{1}\sigma_{2} & \sigma_{2}^2
\end{pmatrix},
$$
in which $\sigma_{1}, \sigma_{2}$ and $\rho$
are the standard deviations of IFMR constants, slopes and the
correlation between them, respectively.
In our analysis, IFMR is parameterised as $M_{\rm final}=\alpha_{0}+\alpha_{1}(M_{\rm initial}-3.0)$,
but many researchers use another form
  $M_{\rm final}=\tilde{\alpha}_{0}+\alpha_{1}M_{\rm initial}$,
  $\tilde{\alpha}_{0}=\alpha_{0}-3\alpha_{1}$, which is the intercept of
the IFMR line with the $M_{\rm initial}=0.0$.
To make our results comparable with others, we transform our results
in the $\tilde{\alpha}_{0}$ (intercept) and $\alpha_{1}$ form.
We denote the population mean of intercept at 0 to be $\tilde{\gamma}_{1}$. 

\begin{table}
\centering
\caption{RMSE and coverage probabilities (CP) of 68.3\% credible intervals (CI) for population-level
 parameters from the Bayesian hierarchical modelling.}
  \vspace{.2cm}
\label{tab:ifmr.pop1}
\begin{tabular}{lllll}
\hline\hline
\multirow{2}{*}{Items}     &    & \multicolumn{3}{c}{Hierarchical} \\\cline{3-5}
                                && RMSE & CP &68.3\% CI \\ \hline
$\gamma_{1}$  && 0.021 & 0.92 & $(0.824,0.972)$ \\ 
$\gamma_{2}$ && 0.009 & 0.68 & $(0.561,0.782)$ \\ 
$\tilde{\gamma}_{1}=\gamma_{1}-3\gamma_{2}$ && 0.034 & 0.76 & $(0.644,0.851)$ \\ 
$\sigma_{1}$  && 0.057 & 0.6 & $(0.48,0.711)$ \\ 
$\sigma_{2}$  && 0.007 & 0.88 & $(0.777,0.945)$ \\ 
$\rho$  && 0.281 & 0.96 & $(0.874,0.993)$ \\ \hline
\end{tabular}
\end{table}

Table \ref{tab:ifmr.pop1} presents the RMSE of point estimates and the actual
 coverage probabilities of 68.3\% credible
interval estimates for population-level parameters using the Bayesian hierarchical approach.
The
68.3\% confidence intervals of the coverage probabilities are computed with 
the Clopper-Pearson exact method \citep{clopper1934use}.
The case-by-case method does not pool star clusters into a population, so it fails
to produce estimates of the population.
From this table, the RMSE of Bayesian hierarchical
 estimates of population-level parameters are small
except $\sigma_{1}$ and $\rho$.
From the actual coverage property of interval estimates, the hierarchical method
tends to produce over-covered interval estimates. For parameters like
$\gamma_{2}, \tilde{\gamma}_{1}$ and $\sigma_{1}$, their interval estimates perform
well, and their actual coverage probabilities are close to the nominal level 68.3\%,
and their 68.3\% confidence interval of coverage contains the nominal value. 
The difficulty of Bayesian hierarchical method in estimating population-level parameters 
is mainly due to two reasons,  the small number of objects, $I = 10$.
Other research have also reported and discussed this problem, refer
to \citet{browne2002introduction, browne2006comparison} for more details.

To further investigate the advantage of the hierarchical analysis in Eq. \eqref{eq:hier},
we take one group of 10 simulated clusters as an example
and compare the estimates of IFMR parameters from the case-by-case and Bayesian hierarchical fits.
After we obtain the MCMC sample for the IFMR parameters, we use its sample means
as point estimates of $\bm{\alpha}_{k}, k=1, \ldots, K$.
We denote the sample means of IFMR parameters from the
cluster-specific fits as $\hat{\bm{\alpha}}_{k}^{\rm CS}$,
and denote those from the hierarchical analysis as $\hat{\bm{\alpha}}_{k}^{\rm Hier}$.

We simultaneously analyse the $10$ simulated datasets using
 a hierarchical model, obtaining a sample of
all parameters from the posterior distribution given in Eq. \ref{eq:jp}. 
In this paper, we focus on estimating
the IFMR parameters $\bm{\alpha}_{1}, \ldots, \bm{\alpha}_{K}$.
After we obtain the MCMC sample for the IFMR parameters, we use its sample means
as point estimates of $\bm{\alpha}_{k}, k=1, \ldots, K$.
We denote the sample means of IFMR parameters from the
cluster-specific analyses as $\hat{\bm{\alpha}}_{k}^{\rm CS}$,
and denote those from the hierarchical analysis as $\hat{\bm{\alpha}}_{k}^{\rm Hier}$.
 
 \begin{table*}
\caption{Point estimates of the IFMR parameters for the $10$ simulated clusters
from both the cluster-specific and hierarchical analyses. Standard errors of estimates
are presented in parentheses.}
\label{tab:ifmr1}
\begin{tabular}{l lll lll ll}
\hline\hline
Cluster  &\multicolumn{2}{c}{True Values} & &\multicolumn{2}{c}{Cluster-specific Estimates}& & \multicolumn{2}{c}{Hierarchical Estimates} \\ \cline{2-3}
\cline{5-6}\cline{8-9}
        & IFMR Const. & IFMR Slope && IFMR Const. & IFMR Slope && IFMR Const. & IFMR Slope \\
($k$) & ($\hat{\alpha}_{k0}$) & ($\hat{\alpha}_{k1}$) && ($\alpha^{\rm CS}_{k0}$) &
($\alpha^{\rm CS}_{k1}$) && ($\hat{\alpha}^{\rm Hier}_{k0}$) & ($\hat{\alpha}^{\rm Hier}_{k1}$)\\\hline
1 & $0.794$ & $0.081$ & & $0.801~(0.018)$ & $0.078~(0.018)$ & & $0.789~(0.024)$ & $0.090~(0.020)$ \\ 
2 & $0.761$ & $0.145$ & & $0.780~(0.039)$ & $0.176~(0.042)$ & & $0.777~(0.032)$ & $0.144~(0.020)$ \\ 
3 & $0.700$ & $0.115$ & & $0.638~(0.048)$ & $0.205~(0.061)$ & & $0.688~(0.036)$ & $0.133~(0.023)$ \\ 
4 & $0.698$ & $0.089$ & & $0.690~(0.030)$ & $0.119~(0.029)$ & & $0.696~(0.026)$ & $0.110~(0.021)$ \\ 
5 & $0.656$ & $0.121$ & & $0.663~(0.040)$ & $0.125~(0.036)$ & & $0.681~(0.030)$ & $0.110~(0.023)$ \\ 
6 & $0.728$ & $0.112$ & & $0.736~(0.020)$ & $0.118~(0.022)$ & & $0.735~(0.019)$ & $0.114~(0.016)$ \\ 
7 & $0.798$ & $0.097$ & & $0.829~(0.047)$ & $0.068~(0.046)$ & & $0.788~(0.030)$ & $0.104~(0.032)$ \\ 
8 & $0.669$ & $0.120$ & & $0.673~(0.020)$ & $0.108~(0.032)$ & & $0.677~(0.017)$ & $0.102~(0.022)$ \\ 
9 & $0.685$ & $0.080$ & & $0.690~(0.049)$ & $0.074~(0.028)$ & & $0.680~(0.037)$ & $0.083~(0.022)$ \\ 
10 & $0.610$ & $0.089$ & & $0.626~(0.036)$ & $0.067~(0.027)$ & & $0.623~(0.031)$ & $0.072~(0.025)$ \\ 
 \hline
\end{tabular}
\end{table*}
\begin{figure}
\centering
\includegraphics[width=0.5\textwidth]{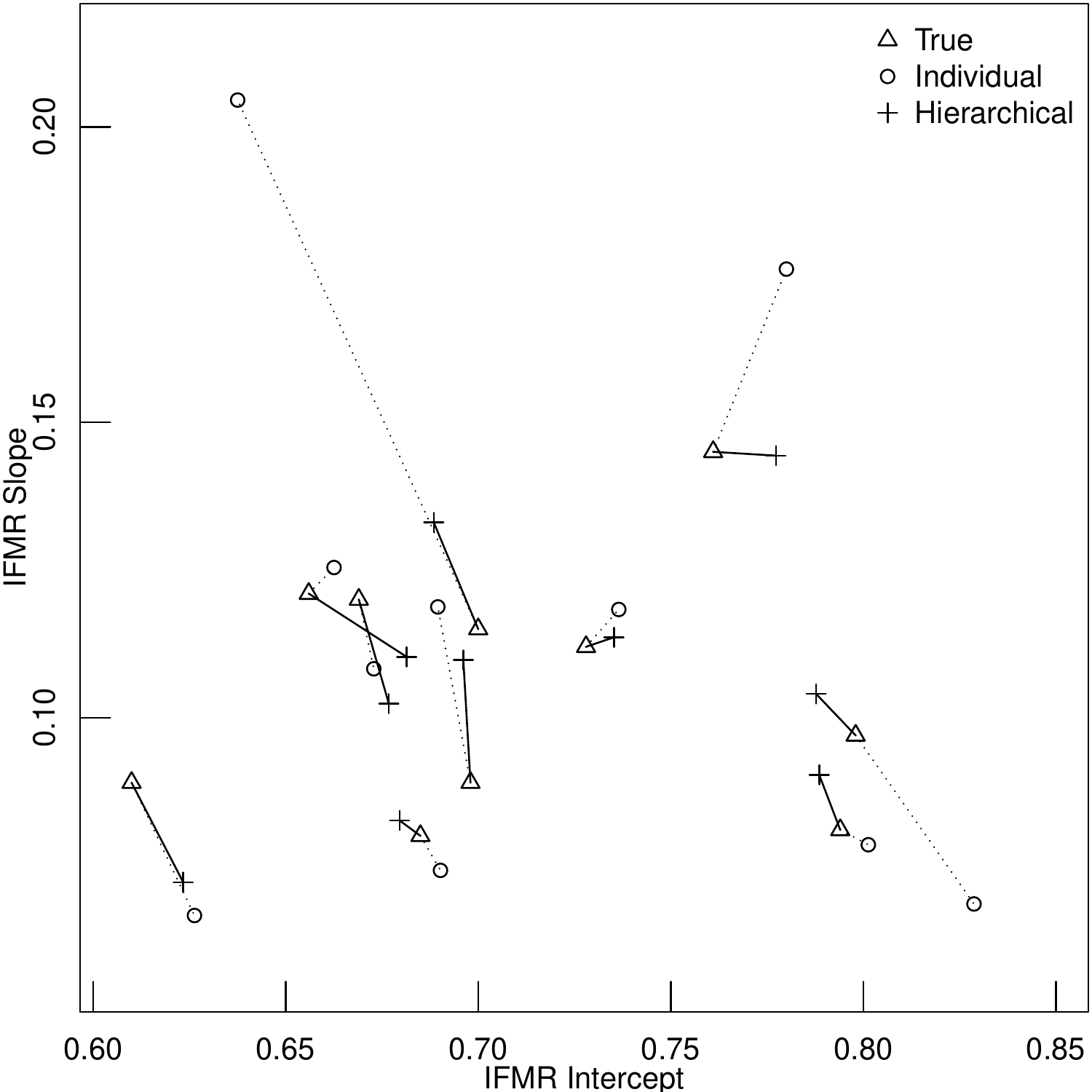}
\caption{Scatter plot of IFMR intercept vs slope. Triangles, circles, and 
plus signs represent true values, cluster-specific and hierarchical estimates
of IFMR parameters for $10$ simulated clusters. Dotted and solid lines
connect the cluster-specific and hierarchical estimates to their true values for
one particular cluster. 
\label{fig:ifmr2}}
\end{figure}
Fig. \ref{fig:ifmr2} presents the scatter plot of IFMR intercept versus slope, with points of
three different shapes (triangle, circle, plus sign) representing true values, cluster-specific (individual)
and hierarchical estimates for IFMR parameters for those $10$ simulated clusters.
To compare the distances between estimates and true values, for one particular cluster,
we connect its true values to cluster-specific and hierarchical estimates with
dotted and solid lines, respectively. From Fig. \ref{fig:ifmr2},
it can be observed that for most clusters the hierarchical model yield more precise estimates of IFMR
parameters than the cluster-specific method. 

Table \ref{tab:ifmr1} reports the true values of the $10$ simulated IFMR parameters and
their estimates from both the hierarchical and cluster-specific analyses.
For most of the clusters, the hierarchical estimates are closer to their true values than are
their cluster-specific counterparts. Specifically, the hierarchical estimates are better for 
clusters 2, 3, 4, 6, 7, 9, and 10. From the aspect of standard errors of estimates in parentheses,
the hierarchical modelling outperforms cluster-specific fits, since it produces
smaller standard errors for every cluster except cluster 1.
Overall, the RMSE of the hierarchical
estimates is $0.012$, i.e., the average deviation of hierarchical estimates from the true values
is $0.012$, while that of the cluster-specific estimates is $0.029$, more than twice the hierarchical value.

\begin{figure}
\centering
\includegraphics[width=0.5\textwidth]{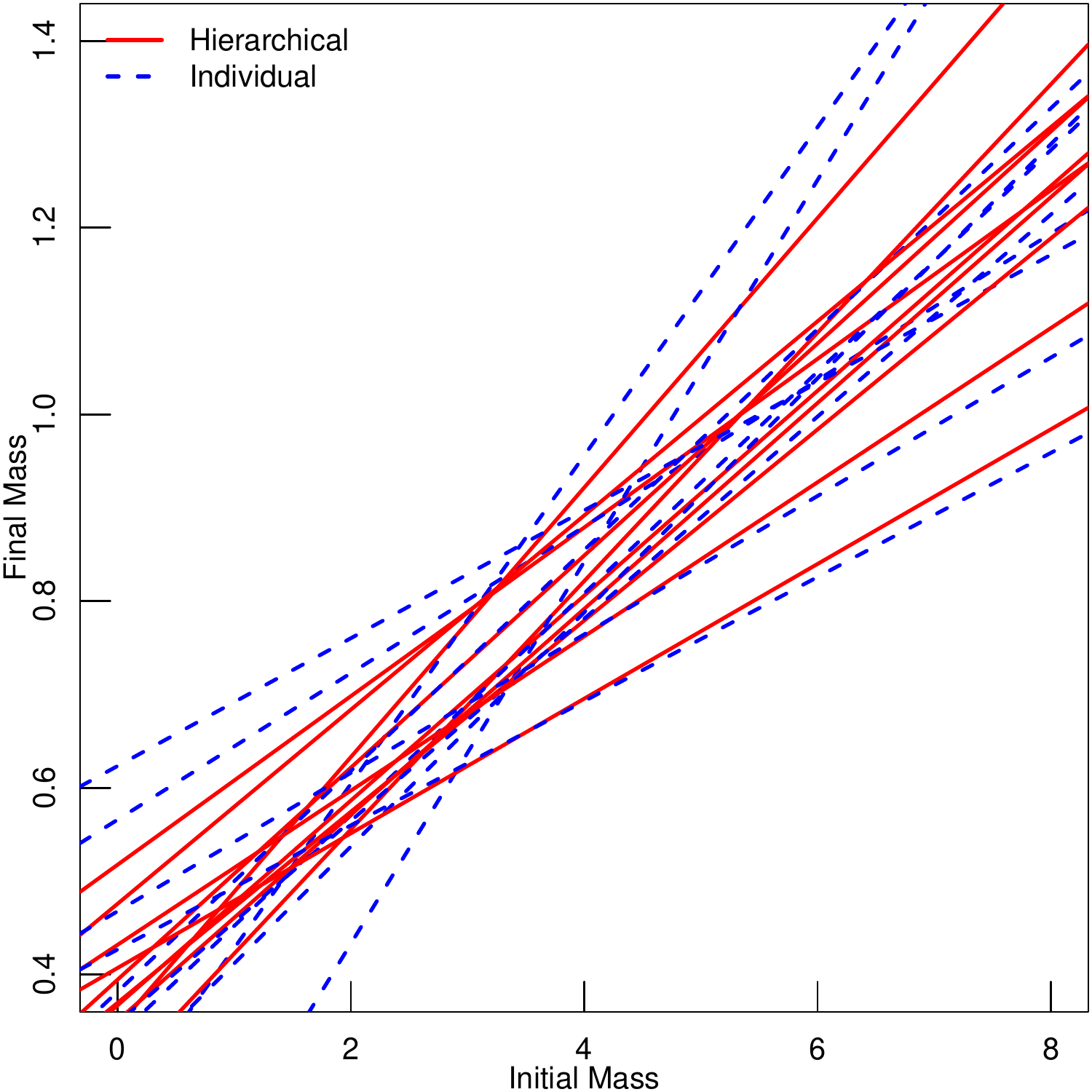}
\caption{Estimated IFMR parameters for $10$ simulated clusters using
both the hierarchical (red solid lines) and cluster-specific (blue dashed lines) analyses.\label{fig:ifmr1}}
\end{figure}

Fig. \ref{fig:ifmr1} illustrates the recovered IFMRs for the $10$ simulated clusters. The red solid lines
are the hierarchical fits and the blue dashed lines are the cluster-specific fits. 
A key feature of the hierarchical estimates is that they tend to cluster toward the centre,
displaying the
shrinkage effect \citep{morris2012shrinkage, gelman2013} of hierarchical models.
Statistically, this property stems from
the assumption that the IFMR parameters of different clusters are generated 
from the same bivariate normal
distribution in Eq. \ref{eq:hier}. Astrophysically, this corresponds to the expectation
that the IFMRs of different clusters, although not identical, are similar. Our Bayesian hierarchical model
is similar in spirit to the method in  
\citet{si2017a}, where we pool ten Galactic halo white dwarfs in a Bayesian hierarchical analysis 
in which we assume that
their ages follow a common normal distribution. \citet{si2017sensitivity} verifies that
even when this normality assumption is violated, estimates based on the
Bayesian hierarchical model still outperform their case-by-case counterparts. 

\begin{table}
  \centering
  \caption{CPU Times of two methods in the IFMR project \label{tab:ifmr.time}}
 \vspace{.2cm}
 \begin{tabular}{l l}
 \hline\hline
Algorithms &  Time \\ \hline
Case-by-case Analysis & About $30$ hours for $10000$ draws\\\hline
 Hierarchical Model & About $30$ hours and $10$ minutes\\
 Two-stage  &  for $10000$ draws\\
 \hline
 \end{tabular}
  \end{table}

Table \ref{tab:ifmr.time} presents the computing time of the case-by-case and Bayesian hierarchical fits.
Because each star cluster may have different number of stars, which affects the computing
time, here is the time for a cluster consisting of $200$ stars, each of which
 has three photometric magnitudes $U, V$, and $I$.
The case-by-case analysis takes about $30$ hours for $10000$ draws and
the hierarchical modelling with two-stage sampler uses the case-by-case fits and costs
additional 10 minutes
 to produce estimates from the hierarchical model.
Throughout this research,
all timings are carried out on a Ubuntu linux server that has
64 AMD Opteron 2.5 GHz processors. We wrote code in R programming language \citep{r2017} to
undertake computations. 
On other computer systems or programming languages, the relative CPU times of
different methods should be similar.

In summary, the Bayesian hierarchical approach with a two-stage sampler produces
shrinkage estimates of the IFMR parameters that have smaller
RMSEs than the case-by-case analysis. 
Also, if the case-by-case samples are available, it only takes 
the two-stage algorithm ten minutes to obtain the estimates 
under the hierarchical model. So we recommend readers to
use our two-stage algorithm to fit hierarchical models.

\section{Data Analysis}\label{sec:data}
 
 \begin{table*}
 \caption{Prior distributions, maximum V magnitudes, and references for the five
 analysed clusters}
 \label{tab:prior}
 \begin{tabular}{llllll}
  \hline\hline
  Cluster & Dist. Mod. &  Metallicity  & Absorption$^a$ & Max. V & Reference\\
  \hline
Hyades & $N(0.0, 0.03^{2})^{b}$ & $N(0.07, 0.05^{2})$  &  $N(0.009, 0.006^{2})$ & 4.5 & \citet{degennaro2009inverting, stein2013}\\
M67 & $N(9.62, 0.091^{2})$ & $N(-0.009, 0.009^{2})$ & $N(0.127, 0.013^{2})$ & 15.0 & \citet{vandenberg2004on,taylor2007the}\\
NGC 188& $N(11.24, 0.1^{2})$ & $N(-0.03, 0.1^{2})$ & $N(0.27, 0.1^{2})$ &15.5 & \citet{von1998wiyn} and\\
&&&&&\citet{meibom2009age}\\
NGC 2168 & $N(10.3, 0.1^{2})$ & $N(-0.2, 0.3^{2})$ & $N(0.682, 0.1^{2})$ & 30.0 & \citet{stein2013}\\
NGC 2477& $N(11.46, 0.0^{2})$ & $N(-0.1, 0.0^{2})$ & $N(0.75, 0.0^{2})$ & 15.5 & \citet{jeffery2011the, stein2013}\\ 
 \hline
 \multicolumn{6}{l}{$^a$ All prior distributions of Absorptions are truncated to positive values.}\\
  \multicolumn{6}{l}{$^b$ The Hyades is analysed with apparent magnitudes converted to absolute magnitudes.}
 \end{tabular}
\end{table*}

In this section we deploy both the cluster-specific and hierarchical analyses 
using photometry for
five star clusters: the Hyades, M67, NGC 188, NGC 2168, and NGC 2477.
In the data analysis, we use \citet{montgomery1999evolutionary} WD interior models
and \citet{bergeron1995} WD atmospheres models. For the MS/RG models, we
use \citet{dotter2008} models for all clusters except NGC2168,
which is too young for the \citet{dotter2008} models, so we
choose \citet{girardi2000evolutionary} models instead.

When BASE-9  fits a star cluster, it uses the MS/RG model to estimate the age and other parameters
of the cluster based on main sequence, main sequence turn-off,
subgiant branch and red giant stars, and uses the WD models to estimate
the ages of the cluster WDs, then it computes the precursor ages for the WDs and uses
the MS/RG models again to determine the initial (ZAMS) masses of the WDs. 
For the \citeauthor{dotter2008} models, the highest mass precursors 
 are $\sim 3.5 M_\odot$ for the metallicity of NGC 2477, and BASE-9 therefore extrapolated 
 the $\log_{10}$(age) versus
precursor mass relation.\label{sen:ms.model}  
This is not
an ideal approach.  Nevertheless, comparing
the \citet{dotter2008} model extrapolation to the \citet{girardi2000evolutionary}
 models yielded similar
results with the  \citeauthor{dotter2008} precursor masses being consistently lower by
just 13.4\% to 17.5\% than the \citeauthor{girardi2000evolutionary} precursor masses.
We return to this point in Section 4.2 when we examine and compare the cluster IFMRs.

\subsection{Cluster-specific Analysis}

We perform the cluster-specific analysis developed by \citet{stein2013},
which uses BASE-9 to deliver MCMC samples of all model parameters
from the respective posterior distribution
for each cluster. 
The prior distributions that we use for distance moduli, metallicities, and absorptions
of the five clusters are shown in Table~\ref{tab:prior}. 
Because the MS/RG models tend to poorly predict the photometry of
faint main sequence stars, we removed main sequence
stars with $V$ magnitudes greater than the cluster-specific thresholds given
in Table \ref{tab:prior}.
Table \ref{tab:prior} also gives the references where we obtain the cluster-specific prior distributions
and cut-off for the $V$ magnitude.
For reading continuity, we present the photometric data and errors for WDs in these five clusters
in Tables \ref{tab:phot}--\ref{tab:phot2} in Appendix \ref{app:phot}.

For NGC 2477 we set
 the prior standard deviations for the distance modulus, absorption, 
 and metallicity to zero.\label{sen:fix.2477} 
The reason for doing this is that NGC 2477
suffers differential reddening, which is not within the BASE-9 model.  
\citeauthor{stein2013} found that by fixing these three cluster parameters at
certain reasonable values consistent with literature estimates,
BASE-9 produces 
good results for the age and IFMR parameters of NGC 2477.
In our analysis, we follow the method of \citet{stein2013}.

Here we elaborate on the prior distribution of distance modulus for the Hyades in Table \ref{tab:prior}. 
\label{sen:hyades.prior}
In the case-by-case (cluster-specific) analysis via BASE-9, we assume that 
all stars in a specific cluster have the same distance modulus.
This assumption is approximately true for clusters fairly far from the Earth. 
For the Hyades, due to the fact that 
its proximity ($\sim 50$ pc) to the Solar System is comparable to
its depth ($\sim 10-20$ pc),
 its member stars have significantly different distances, which
 violates the equal distance assumption in the BASE-9. \label{sen:hyades.dist}
 To address
 this problem, \citet{degennaro2009white} adjusted the magnitudes of each star for its
 distance using the precise distance estimates obtained
by \citet{bruijne2001a}.
 Each Hyades star was offset to a nominal distance modulus of $m - M_{V} = 0.0$, i.e.,
 10 pc. We therefore set the prior distance modulus to be a Gaussian distribution
 with mean $0$. Additionally, because the Hyades is well-studied and the uncertainty
of its distance modulus is small, we take $0.03$ as the prior standard deviation.
After we obtain the MCMC samples for the Hyades from the case-by-case analysis,
we add the average distance modulus from multiple studies, $3.40$, \citep{perryman1998the,degennaro2009white}
 to the MCMC sample
of distance modulus and thereby recover the posterior sample of distance modulus for Hyades
with BASE-9. For details, refer to
\citet{degennaro2009white, degennaro2009inverting, stein2013}.

\begin{figure*}
\centering
\includegraphics[width=0.9\textwidth]{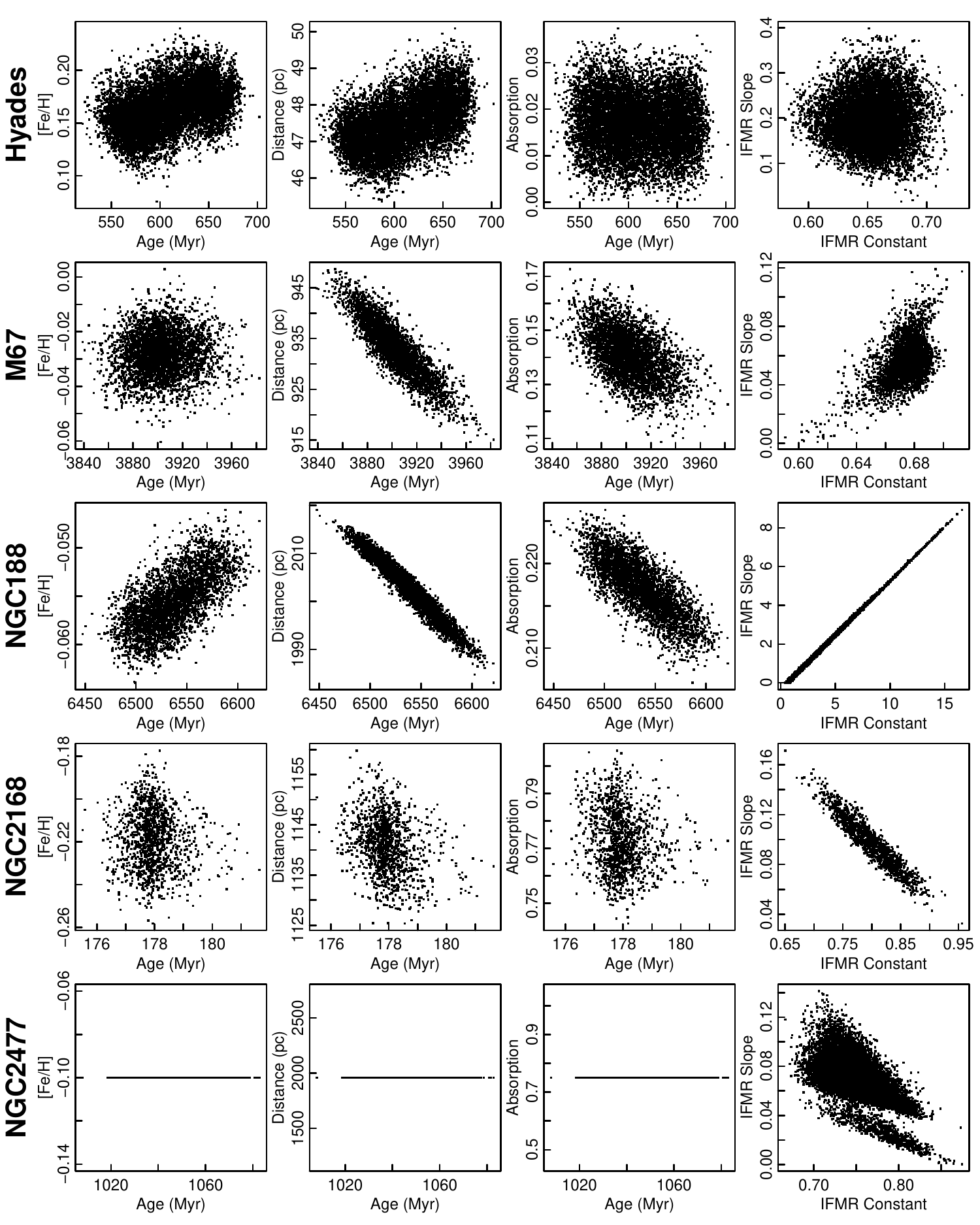}
\caption{Cluster-specific results: projections of the joint posterior distributions onto
the two dimensional planes of (from left to right) age--metallicity, age--distance, age--absorption, and
IFMR intercept--IFMR slope for the five analysed clusters.\label{fig:clusters}}
\end{figure*}

The MCMC samples from the cluster-specific analyses appear in 
 Fig. \ref{fig:clusters}. Each row corresponds to one cluster, and the columns
provide scatter plots of various parameter combinations.  
Because the prior standard deviations of metallicity,
distance modulus, and absorption are set to zero for NGC 2477, the scatter
plots of age--metallicity, age--distance, age--absorption degenerate into lines.
The scatter plot of the IFMR parameters for NGC 2477 has two separate modes.
The upper mode, accounting for 90.44\% of the distribution,
 tends to have a larger IFMR slope than the lower one, constituting 9.56\% of the
posterior distribution.
The most likely explanation for the bimodal nature is uncertainty in cluster 
membership of one or more stars.

The rightmost column of Fig. \ref{fig:clusters} displays the scatter plots of the IFMR parameters for
the five clusters under the cluster-specific analyses.
For all of the clusters except NGC 188, the range of the IFMR intercept is $0.55$ to $0.95$
and that of the IFMR slope is $0.0$ to $0.4$, which are both 
quite consistent with the results in \citet{salaris2009}
and \citet{williams2009}.
However, the IFMR parameters for NGC 188 are both
 far from their commonly accepted ranges. 
This appears to be because all of the WDs in NGC 188 
have similar ZAMS masses ($1.17$ to $1.24~M_{\sun}$),
but their WD masses vary significantly ($0.52$ to $0.80~M_{\sun}$).
We do not address these particular properties of NGC 188 WDs,
instead leaving them to be discussed in Section \ref{sec:hier.ana}. This difference
in the fitted IFMRs provides an opportunity
to test the power of the hierarchical model.
In the next section, we simultaneously analyse the five clusters
with a hierarchical model. This allows us to borrow strength among the
 clusters and provides more reliable estimates of the
IFMR parameters, particularly for NGC 188.

The estimates of cluster parameters from the cluster-specific analyses
are shown in the lower part in Table \ref{tab:clus1}.
The IFMR parameters -- intercept and slope -- are in the last two columns. \label{sen:ifmr.diff}
From the cluster-specific analyses, the estimates of the IFMR parameters vary significantly 
from cluster to cluster. Most noticeably, the IFMR estimates of NGC 188 are unrealistic with very large
standard errors. The other star clusters also exhibit significant differences in their estimated IFMR
parameters, especially in the IFMR slopes. We do not know the exact reasons for these divergences.
One possible explanation is that we assume each star cluster has its own linear IFMR, which
affects the estimates of the IFMR parameters. Yet many researchers argue that the IFMR is nonlinear 
 \citep{marigo2007evolution,meng2008initial,choi2016mesa}.   
Alternatively cluster metallicity may affect the 
IFMR \citep{meng2008initial,zhao2012the}.
The metallicities of these five clusters vary significantly, which might cause the
divergences of their IFMR parameters.

Here we investigate the sensitivity of the cluster's IFMR parameters to its WD mass range.
We still assume the linear functional form of the IFMRs and use NGC 2477 as an example.
NGC 2477 has seven WDs in the original cluster-specific analysis and among them three have
ZAMS between 2 to 4 $M_{\odot}$, with the other four above $4 M_{\odot}$.
In this test we remove the three low mass WDs below $4 M_{\odot}$ from NGC 2477,
 use BASE-9 to fit the modified dataset and compare the fitted IFMR parameters.
\begin{figure*}
\centering
\includegraphics[width=0.9\textwidth]{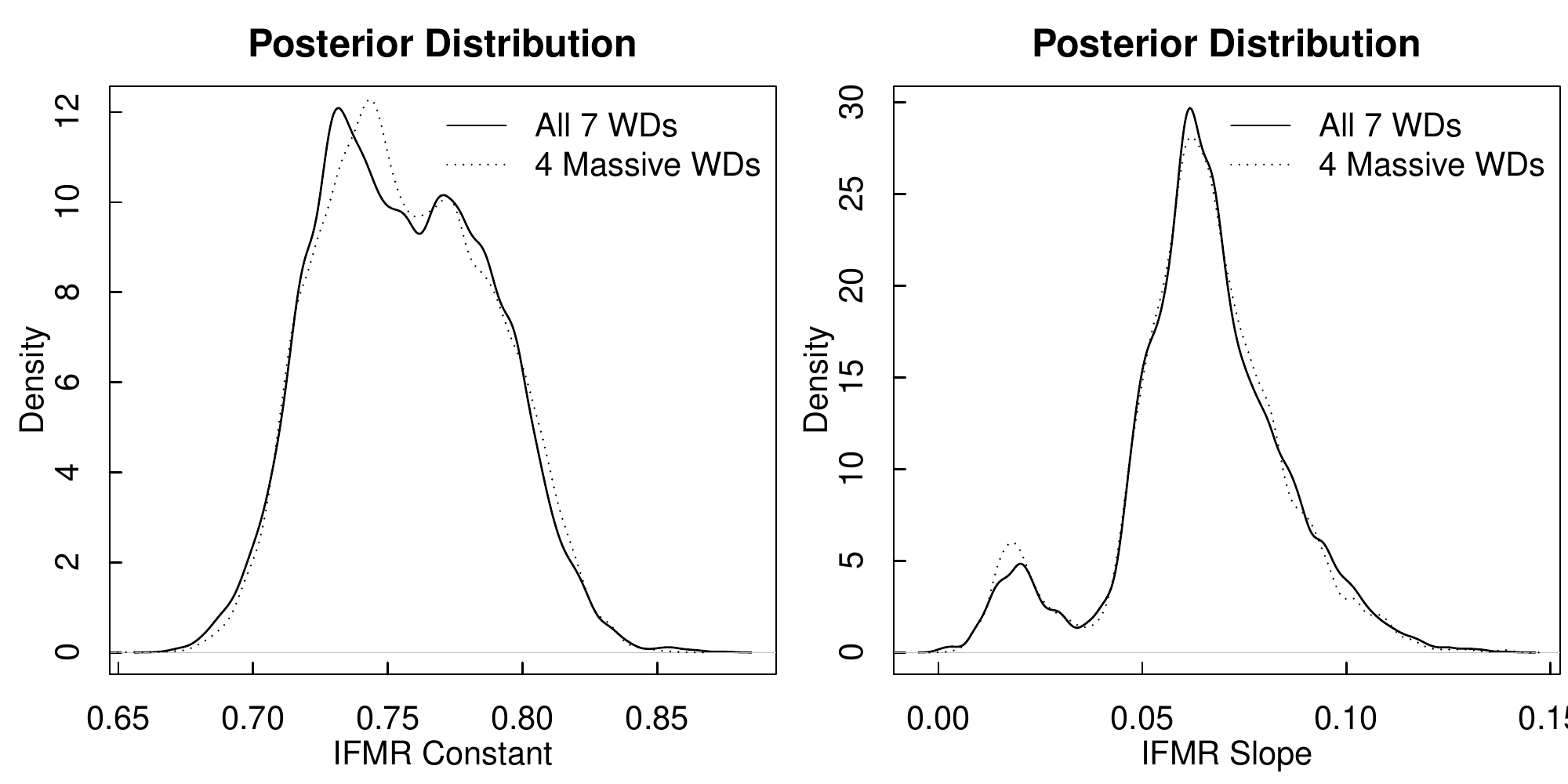}
\caption{Two fits of NGC 2477: solid line is from the original fit (including all WDs in NGC 2477)
 and dotted line is from the fit after
three less massive white dwarfs are removed.}
\label{fig:ifmr.range}
\end{figure*}
Fig. \ref{fig:ifmr.range} displays the histograms of IFMR parameters for NGC 2477 under the two conditions:
including all seven WDs and only including the four WDs above $4 M_{\odot}$. \label{sen:ifmr.range}
The left and right panels show the posterior distributions of IFMR constant and slope for NGC 2477, respectively.
The solid and dotted histograms represent results from the cases: 1) all seven WDs in NGC 2477 are used and
2) only the WDs above $4 M_{\odot}$ are used, respectively.
The histograms of both IFMR constant and slope remain essentially the same in both
cases (with tiny difference caused
by simulation errors), which means that
the estimates of IFMR parameters for NGC 2477 vary insignificantly even as its initial mass 
range diminishes. This small experiment implies that at least for the current NGC 2477 data, that the 
progenitor mass range does not affect the IFMR parameters.

Studies have shown that
 the metallicity may affect the IFMR parameters of a cluster
\citep[e.g.,][]{kalirai2005the,catalan2008the,meng2008initial,zhao2012the}.
 We have explored the quantity and quality of data required to test whether the
IFMR intercept and slope depends on metallicity.
We can investigate the effect of metallicity on the IFMR parameters via an extension
of our Bayesian hierarchical
model. To achieve this, we adjust the bivariate Gaussian assumption
on IFMR parameters $\bm{\alpha}_{k}$ in Eq. \ref{eq:hier} to be
$$\bm{\alpha}_{k} =
\begin{pmatrix} \alpha_{k0} \\ \alpha_{k1} \end{pmatrix} \sim
\bm{N}\Big(\bm{B}\bm{\eta}_{k}, \bm{\Gamma}\Big),
$$
with 
$$\bm{B} = \begin{pmatrix} b_{11} & b_{12}\\ b_{21} & b_{22} \end{pmatrix}
, \bm{\eta}_{k} = \begin{pmatrix} 1 \\ \theta_{{\rm[Fe/H]},k} \end{pmatrix},$$
where matrix $\bm{B}$ is the effect of metallicity $\theta_{{\rm[Fe/H]},k}$ of cluster $k$ on
its IFMR parameters $\bm{\alpha}_{k}$. 
Our two-stage algorithm has the capacity to fit this complicated hierarchical model.
This model introduces four more parameters in the effect matrix $\bm{B}$, yet at present 
we only have five clusters in this study. So for the present study we maintain the simple model of
Eq. \eqref{eq:hier} and we plan to investigate the effect of metallicity on
the IFMR once we have a sufficient number of stellar clusters.

\subsection{Hierarchical Analysis}\label{sec:hier.ana}

\begin{table*}
\centering
\small
\caption{Parameter estimates for the five clusters under both the hierarchical and cluster-specific fits}
\label{tab:clus1}
\begin{tabular}{llllllll}
\hline\hline
&Cluster & $\log_{10}$(Age) & $m-M_{V}$ & [Fe/H]   & Absorption  & IFMR Intercept & IFMR Slope \\ \hline
 \multirow{6}{*}{\begin{tabular}[l]{@{}l@{}}Hierarchical\\ Estimates\end{tabular}}& Hyades & $8.773\pm0.026$ & $-0.000\pm0.030$ & $0.157\pm0.020$ & $0.017\pm0.006$ & $0.660\pm0.020$ & $0.140\pm0.050$ \\ 
 & M67 & $9.591\pm0.002$ & $9.850\pm0.010$ & $-0.029\pm0.007$ & $0.142\pm0.008$ & $0.680\pm0.010$ & $0.060\pm0.010$ \\ 
& NGC188 & $9.815\pm0.002$ & $11.510\pm0.010$ & $-0.056\pm0.003$ & $0.218\pm0.003$ & $0.750\pm0.120$ & $0.090\pm0.050$ \\ 
& NGC2168 & $8.250\pm0.001$ & $10.290\pm0.010$ & $-0.219\pm0.015$ & $0.774\pm0.012$ & $0.790\pm0.040$ & $0.100\pm0.020$ \\ 
& NGC2477 & $9.019\pm0.004$ & $11.460\pm0.000$ & $-0.100\pm0.000$ & $0.750\pm0.000$ & $0.750\pm0.030$ & $0.070\pm0.020$ \\  \hline 
 \multirow{6}{*}{\begin{tabular}[l]{@{}l@{}}Cluster-specific\\ Estimates\end{tabular}}& Hyades & $8.785\pm0.028$ & $0.010\pm0.030$ & $0.164\pm0.021$ & $0.017\pm0.006$ & $0.650\pm0.020$ & $0.200\pm0.060$ \\ 
& M67 & $9.591\pm0.002$ & $9.850\pm0.010$ & $-0.029\pm0.007$ & $0.142\pm0.008$ & $0.680\pm0.010$ & $0.050\pm0.010$ \\ 
& NGC188 & $9.815\pm0.002$ & $11.510\pm0.010$ & $-0.056\pm0.003$ & $0.217\pm0.003$ & $4.520\pm3.130$ & $2.200\pm1.750$ \\ 
& NGC2168 & $8.250\pm0.001$ & $10.290\pm0.010$ & $-0.221\pm0.015$ & $0.775\pm0.012$ & $0.810\pm0.050$ & $0.100\pm0.020$ \\ 
& NGC2477 & $9.019\pm0.004$ & $11.460\pm0.000$ & $-0.100\pm0.000$ & $0.750\pm0.000$ & $0.760\pm0.030$ & $0.070\pm0.020$ \\ \hline
\end{tabular}
\end{table*}

\begin{figure*}
\centering
\includegraphics[width=0.8\textwidth]{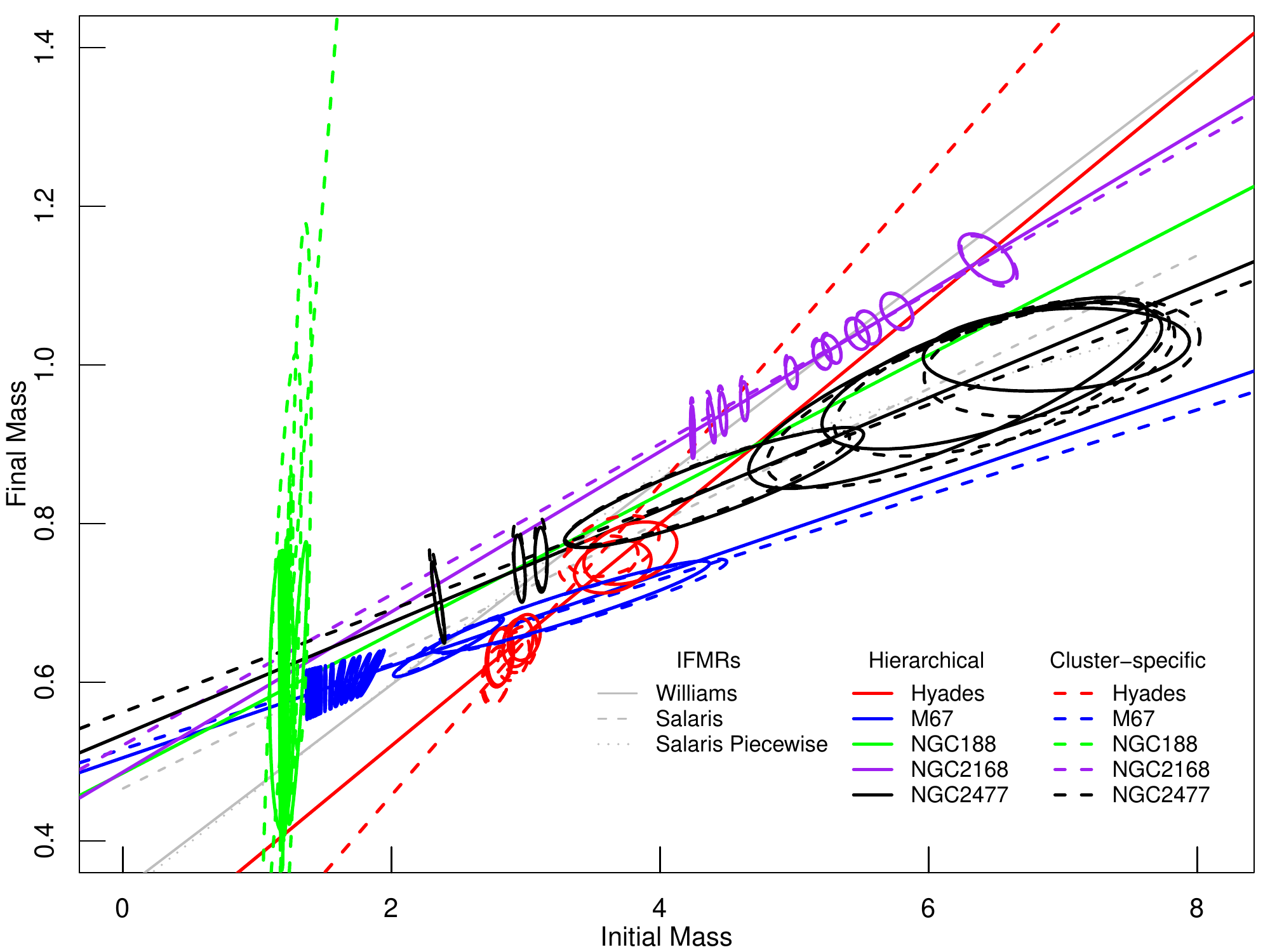}
\caption{The estimated IFMRs for the five clusters obtained with both hierarchical
and cluster-specific analyses. The ovals are 68.3\% contours of the joint posterior distribution of
the ZAMS and WD masses
 for the WDs in each cluster.
 The colour scheme of red, blue, green, purple, and black corresponds to the Hyades, M67, 
 NGC 188, NGC 2168, and NGC 2477, respectively. 
 The Williams, Salaris and Salaris piecewise IFMRs are plotted as solid, dashed, and dotted grey lines. 
 \label{fig:compound}}
\end{figure*}

In this section, we present the result obtained 
under the hierarchical analysis in Section \ref{sec:hier1}.  
We deploy the two-stage algorithm in Appendix \ref{app:alg} to obtain the MCMC samples for 
all model parameters. For simplicity, we compute posterior sample means and standard deviations to
summarise the posterior distributions of each parameter.
Table \ref{tab:clus1} compares the estimates and error bars for all five clusters
obtained using the hierarchical and cluster-specific methods.  In all five cases,
the estimates of $\log_{10}$(Age), $m-M_{V}$, [Fe/H] and absorption are 
nearly the same for the hierarchical and cluster-specific fits.
The estimates of the IFMR parameters for NGC 188, however, differ substantially. From the cluster-specific
analysis, the posterior mean of the IFMR for NGC 188 is $M_{\rm WD}=4.522+2.195(M_{\rm ZAMS}-3.0)$.
This implies that a star with ZAMS mass $3~M_{\sun}$ has a WD mass of $4.522~M_{\sun}$ and 
that WD mass increases by $2.195~M_{\sun}$ for each additional $M_{\sun}$ in its ZAMS mass.
Clearly, this result is nonsense: it violates conservation of mass. The reason
the cluster-specific analysis results in this
bizarre IFMR is that the ZAMS masses of WDs in NGC 188
are in a narrow range, $1.17$ to $1.33~M_{\sun}$, so they fail to constrain the IFMR parameters
over the whole ZAMS mass range.  
By contrast, the hierarchical model yields reasonable estimates for the IFMR parameters of
NGC 188, $M_{\rm WD}=0.749+0.088(M_{\rm ZAMS}-3.0)$.
For the other clusters, the hierarchical and cluster-specific estimates have slight differences
due to the shrinkage effects of the hierarchical model, which are further
illustrated in Fig. \ref{fig:compound}. The hierarchical and
cluster-specific estimates of the IFMR slopes differ by about one standard deviation
for both the Hyades and M67. 
This is caused by the shrinkage effect: IFMR slopes of the five clusters
shrink their grand mean. M67 has the shallowest IFMR and
the Hyades has the second steepest IFMR, shallower only than NGC 188, so they are more 
substantially affected
by the hierarchical analysis.

\begin{figure}
\centering
\includegraphics[width=0.5\textwidth]{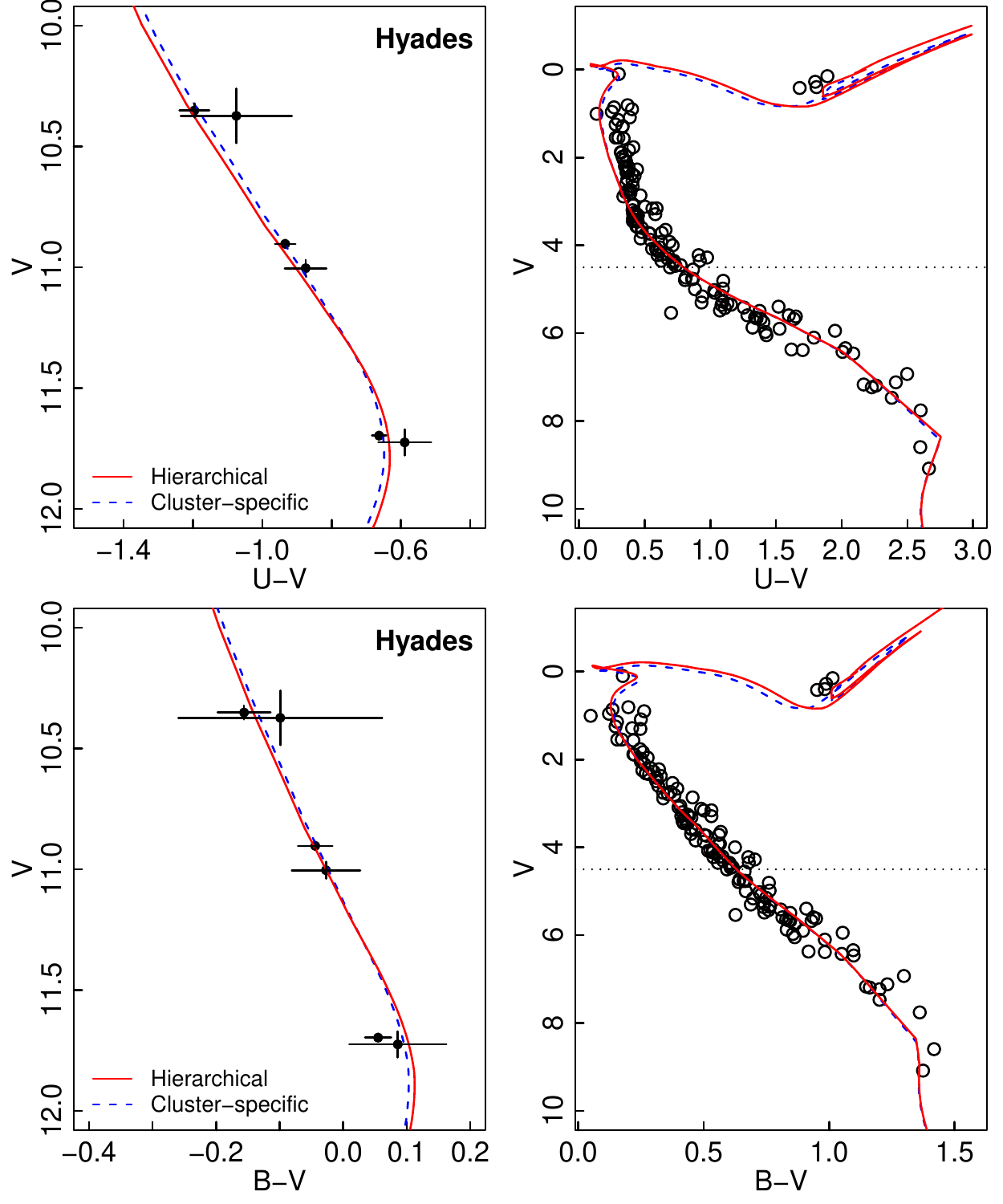}
\caption{Color magnitude diagrams for the Hyades based on both the hierarchical and cluster-specific estimates.
The solid (red) lines are from the hierarchical fits and dashed (blue) lines are from the cluster-specific fits.
\label{fig:cmd1}}
\end{figure}

\begin{figure}
\centering
\includegraphics[width=0.5\textwidth]{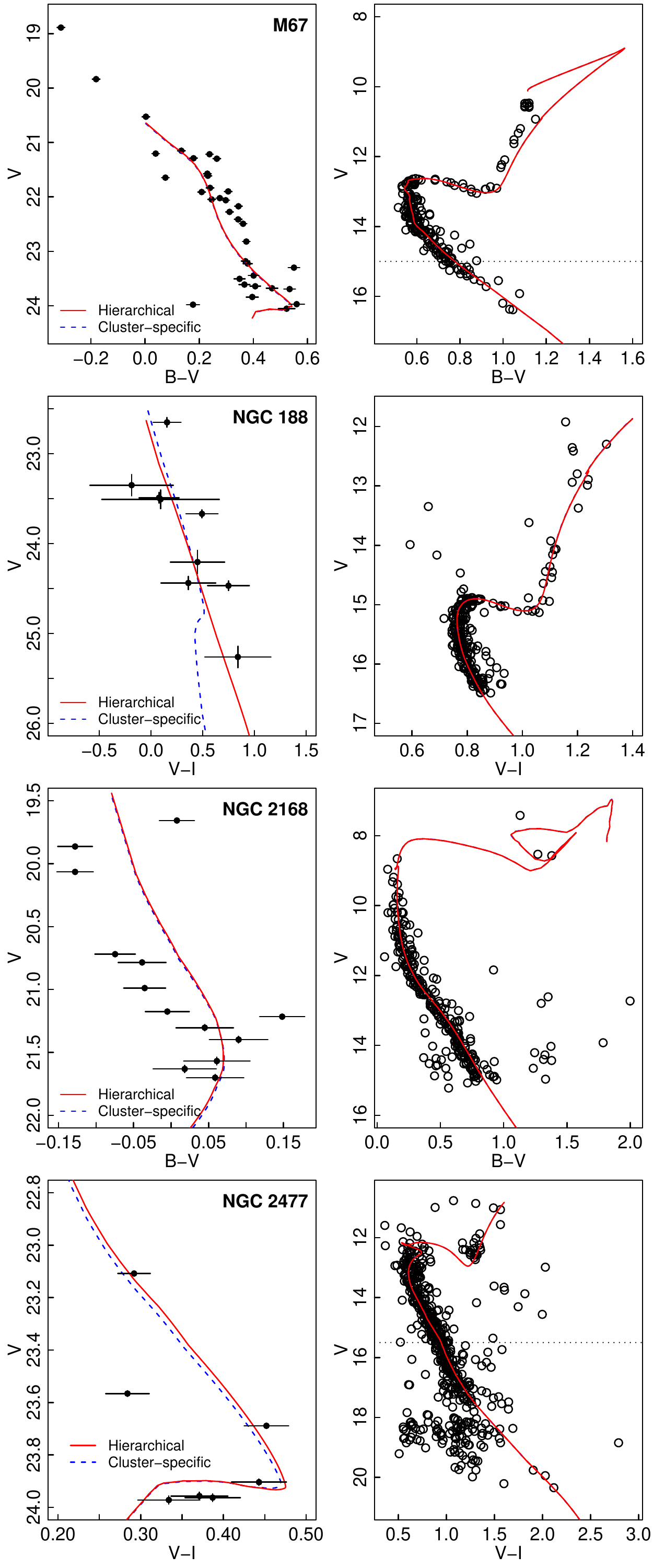}
\caption{Color magnitude diagrams for (from top to bottom) M67, NGC 188,
NGC 2168, NGC 2477, based on both the hierarchical and cluster-specific analyses.
The solid (red) lines indicate the hierarchical fits and dashed (blue) lines indicate the cluster-specific fits. 
\label{fig:cmd2}}
\end{figure}

Figs. \ref{fig:cmd1} and \ref{fig:cmd2} plot the colour magnitude diagrams (CMD) for the five clusters.
Fig. \ref{fig:cmd1} presents the U-V and B-V CMDs for the Hyades and the rows of Fig. \ref{fig:cmd2}
display CMDs of clusters (from top to bottom) M67, NGC 188, NGC 2168, and NGC 2477.
The solid (red) lines are from the hierarchical fits and the dashed (blue)
lines are from the cluster-specific fits. In each row of Figs. \ref{fig:cmd1} and \ref{fig:cmd2}, 
the left panel displays a close up of the WD region of the CMD
 and the right panel shows the MS/RG stars. 
The CMDs for the MS/RG regions from both the hierarchical and cluster-specific fits are similar for
all five clusters. Likewise, the CMDs for the WDs are also similar,
 for all clusters except NGC 188.
For NGC 188, the cluster-specific CMD (dashed blue)
is quite far from the dimmest WD, while the hierarchical CMD
(solid red) is consistent with all of the cluster's WDs. This illustrates
an advantage of the hierarchical model.

Fig. \ref{fig:compound} compares the estimated IFMRs (plotted as lines) for the five clusters 
along with the 68.3\% contours (plotted as ovals) of the joint posterior distribution of
 initial (ZAMS) and final (WD) masses for each WD in each cluster. Results for both 
 the hierarchical (solid)
and cluster-specific (thick dashed) analyses are plotted.    
Colours (red, blue, green, purple, and black) correspond to five clusters
(Hyades, M67, NGC 188, NGC 2168 and NGC 2477, respectively).
The solid IFMR lines from the hierarchical fits tend to
be in the centre and are consistent with the
with prior IFMRs (e.g. Williams, Bolte \& Koester, 2004; Salaris et al. 2009; plotted as grey
 solid, dashed and dotted lines, 
respectively.
The thick dashed fitted IFMRs from the cluster-specific approach have more uncertainty.
The most striking feature of this figure is that the cluster-specific analysis of NGC 188 yields an
unreasonably steep IFMR (plotted as a dashed green line), 
whereas the hierarchical model produces a much
shallower and more reasonable IFMR for NGC 188 (plotted as a solid green line).

Tables \ref{tab:mass1} --- \ref{tab:mass3} 
present the initial masses, final masses, and membership probabilities for WDs in the five
clusters, based on both the cluster-specific and hierarchical modelling approaches.
%%%%%%%
\begin{table*}
\caption{Results from both the hierarchical and cluster-specific analyses for the WDs in the Hyades}
\label{tab:mass1}
\begin{tabular}{ll lll l lll}
\hline\hline
Cluster & WD & \multicolumn{3}{c}{Hierarchical Estimates} &&\multicolumn{3}{c}{Cluster-specific Estimates}\\ 
\cline{3-5}\cline{7-9}
& & ZAMS Mass & WD Mass & Mem. Prob. & &  ZAMS Mass & WD Mass & Mem. Prob. \\ \hline
Hyades& HZ14 & $2.801\pm0.064$ & $0.631\pm0.024$ & $1.000$ & & $2.776\pm0.067$ & $0.607\pm0.024$ & $1.000$ \\ 
& VR16 & $2.813\pm0.064$ & $0.632\pm0.024$ & $1.000$ & & $2.787\pm0.067$ & $0.609\pm0.024$ & $1.000$ \\  
& HZ7 & $2.942\pm0.074$ & $0.650\pm0.020$ & $1.000$ & & $2.910\pm0.078$ & $0.633\pm0.020$ & $1.000$ \\ 
& VR7 & $2.988\pm0.079$ & $0.657\pm0.019$ & $1.000$ & & $2.954\pm0.084$ & $0.641\pm0.019$ & $1.000$ \\
& HZ4 & $3.779\pm0.229$ & $0.763\pm0.026$ & $1.000$ & & $3.649\pm0.228$ & $0.772\pm0.025$ & $1.000$ \\ 
& LB227 & $3.648\pm0.188$ & $0.746\pm0.022$ & $1.000$ & & $3.543\pm0.194$ & $0.752\pm0.021$ & $1.000$ \\ \hline
\end{tabular}
\end{table*}

\begin{table*}
\caption{Results from the Hierarchical and Cluster-specific fits on WDs in M67}
\label{tab:mass2}
\begin{tabular}{ll lll l lll}
\hline\hline
Cluster & WD & \multicolumn{3}{c}{Hierarchical Estimates} &&\multicolumn{3}{c}{Cluster-specific Estimates}\\ 
\cline{3-5}\cline{7-9}
& & ZAMS Mass & WD Mass & Mem. Prob. & &  ZAMS Mass & WD Mass & Mem. Prob. \\ \hline
M67& WD 1 & $1.615\pm0.021$ & $0.598\pm0.019$ & $1.000$ & & $1.619\pm0.021$ & $0.602\pm0.017$ & $1.000$ \\ 
& WD 2 & $1.438\pm0.003$ & $0.588\pm0.020$ & $1.000$ & & $1.438\pm0.003$ & $0.593\pm0.018$ & $1.000$ \\ 
& WD 3 & $1.429\pm0.003$ & $0.588\pm0.020$ & $1.000$ & & $1.429\pm0.003$ & $0.592\pm0.018$ & $1.000$ \\ 
& WD 4 & $1.412\pm0.003$ & $0.587\pm0.020$ & $0.000$ & & $1.412\pm0.003$ & $0.591\pm0.018$ & $0.000$ \\ 
& WD 5 & $1.364\pm0.003$ & $0.584\pm0.021$ & $0.209$ & & $1.364\pm0.003$ & $0.589\pm0.019$ & $0.191$ \\ 
& WD 6 & $1.426\pm0.003$ & $0.587\pm0.020$ & $1.000$ & & $1.426\pm0.003$ & $0.592\pm0.018$ & $1.000$ \\ 
& WD 7 & $1.414\pm0.003$ & $0.587\pm0.020$ & $1.000$ & & $1.414\pm0.003$ & $0.591\pm0.018$ & $1.000$ \\ 
& WD 8 & $1.396\pm0.003$ & $0.586\pm0.020$ & $1.000$ & & $1.395\pm0.003$ & $0.590\pm0.018$ & $1.000$ \\ 
& WD 9 & $1.389\pm0.003$ & $0.585\pm0.020$ & $1.000$ & & $1.388\pm0.003$ & $0.590\pm0.018$ & $1.000$ \\ 
& WD 10 & $1.556\pm0.006$ & $0.595\pm0.019$ & $1.000$ & & $1.558\pm0.006$ & $0.599\pm0.017$ & $1.000$ \\ 
& WD 11 & $1.431\pm0.003$ & $0.588\pm0.020$ & $0.983$ & & $1.431\pm0.003$ & $0.592\pm0.018$ & $0.981$ \\ 
& WD 12 & $1.394\pm0.003$ & $0.586\pm0.020$ & $0.329$ & & $1.393\pm0.003$ & $0.590\pm0.018$ & $0.253$ \\ 
& WD 13 & $1.370\pm0.003$ & $0.584\pm0.021$ & $1.000$ & & $1.370\pm0.003$ & $0.589\pm0.018$ & $1.000$ \\ 
& WD 14 & $1.506\pm0.003$ & $0.592\pm0.019$ & $0.999$ & & $1.507\pm0.003$ & $0.596\pm0.017$ & $0.999$ \\ 
& WD 15 & $1.454\pm0.003$ & $0.589\pm0.020$ & $1.000$ & & $1.454\pm0.003$ & $0.594\pm0.018$ & $1.000$ \\ 
& WD 16 & $1.828\pm0.056$ & $0.610\pm0.019$ & $1.000$ & & $1.838\pm0.051$ & $0.614\pm0.017$ & $1.000$ \\ 
& WD 17 & $1.440\pm0.003$ & $0.588\pm0.020$ & $1.000$ & & $1.440\pm0.003$ & $0.593\pm0.018$ & $1.000$ \\ 
& WD 18 & $1.473\pm0.003$ & $0.590\pm0.019$ & $0.595$ & & $1.473\pm0.003$ & $0.595\pm0.017$ & $0.544$ \\ 
& WD 19 & $1.549\pm0.005$ & $0.594\pm0.019$ & $1.000$ & & $1.550\pm0.006$ & $0.599\pm0.017$ & $1.000$ \\ 
& WD 20 & $1.389\pm0.003$ & $0.585\pm0.020$ & $0.173$ & & $1.388\pm0.003$ & $0.590\pm0.018$ & $0.221$ \\ 
& WD 21 & $1.363\pm0.003$ & $0.584\pm0.021$ & $0.033$ & & $1.363\pm0.003$ & $0.589\pm0.019$ & $0.030$ \\ 
& WD 22 & $2.424\pm0.270$ & $0.644\pm0.025$ & $1.000$ & & $2.424\pm0.252$ & $0.645\pm0.023$ & $1.000$ \\ 
& WD 23 & $3.335\pm0.682$ & $0.695\pm0.038$ & $1.000$ & & $3.365\pm0.747$ & $0.695\pm0.040$ & $1.000$ \\ 
& WD 24 & $1.398\pm0.003$ & $0.586\pm0.020$ & $0.092$ & & $1.398\pm0.003$ & $0.591\pm0.018$ & $0.056$ \\ 
& WD 25 & $1.671\pm0.030$ & $0.601\pm0.019$ & $0.999$ & & $1.677\pm0.029$ & $0.605\pm0.017$ & $1.000$ \\ 
& WD 26 & $1.768\pm0.047$ & $0.607\pm0.019$ & $1.000$ & & $1.777\pm0.044$ & $0.610\pm0.017$ & $1.000$ \\ 
& WD 27 & $1.466\pm0.003$ & $0.590\pm0.020$ & $0.976$ & & $1.466\pm0.003$ & $0.594\pm0.018$ & $0.971$ \\ 
& WD 28 & $1.708\pm0.036$ & $0.603\pm0.019$ & $1.000$ & & $1.715\pm0.034$ & $0.607\pm0.017$ & $1.000$ \\ 
& WD 29 & $1.607\pm0.019$ & $0.598\pm0.019$ & $1.000$ & & $1.611\pm0.019$ & $0.602\pm0.017$ & $1.000$ \\ 
& WD 30 & $1.812\pm0.055$ & $0.609\pm0.019$ & $0.000$ & & $1.822\pm0.050$ & $0.613\pm0.017$ & $0.000$ \\ 
& WD 31 & $1.437\pm0.003$ & $0.588\pm0.020$ & $1.000$ & & $1.437\pm0.003$ & $0.593\pm0.018$ & $1.000$ \\ 
& WD 32 & $1.856\pm0.058$ & $0.612\pm0.019$ & $0.996$ & & $1.867\pm0.052$ & $0.615\pm0.017$ & $0.998$ \\ 
& WD 33 & $1.412\pm0.003$ & $0.587\pm0.020$ & $1.000$ & & $1.412\pm0.003$ & $0.591\pm0.018$ & $1.000$ \\ 
& WD 34 & $1.449\pm0.003$ & $0.589\pm0.020$ & $0.098$ & & $1.449\pm0.003$ & $0.593\pm0.018$ & $0.073$ \\ 
& WD 35 & $1.599\pm0.016$ & $0.597\pm0.019$ & $0.093$ & & $1.602\pm0.017$ & $0.601\pm0.017$ & $0.060$ \\  \hline
\end{tabular}
\end{table*}

\begin{table*}
\caption{Results from the hierarchical and cluster-specific analyses for the WDs in NGC188, NGC2168 and NGC2477}
\label{tab:mass3}
\begin{tabular}{ll lll l lll}
\hline\hline
Cluster & WD & \multicolumn{3}{c}{Hierarchical Estimates} &&\multicolumn{3}{c}{Cluster-specific Estimates}\\ 
\cline{3-5}\cline{7-9}
& & ZAMS Mass & WD Mass & Mem. Prob. & &  ZAMS Mass & WD Mass & Mem. Prob. \\ \hline
NGC188& WD 1 & $1.169\pm0.003$ & $0.588\pm0.117$ & $1.000$ & & $1.171\pm0.003$ & $0.510\pm0.128$ & $1.000$ \\ 
& WD 2 & $1.184\pm0.003$ & $0.590\pm0.116$ & $1.000$ & & $1.185\pm0.003$ & $0.538\pm0.122$ & $1.000$ \\ 
& WD 3 & $1.187\pm0.002$ & $0.590\pm0.116$ & $1.000$ & & $1.188\pm0.002$ & $0.544\pm0.121$ & $1.000$ \\ 
& WD 4 & $1.184\pm0.060$ & $0.590\pm0.117$ & $1.000$ & & $1.177\pm0.100$ & $0.519\pm0.326$ & $1.000$ \\ 
& WD 5 & $1.191\pm0.002$ & $0.590\pm0.116$ & $1.000$ & & $1.191\pm0.002$ & $0.551\pm0.120$ & $1.000$ \\ 
& WD 6 & $1.208\pm0.006$ & $0.592\pm0.116$ & $1.000$ & & $1.210\pm0.010$ & $0.595\pm0.130$ & $1.000$ \\ 
& WD 7 & $1.219\pm0.009$ & $0.592\pm0.116$ & $1.000$ & & $1.226\pm0.017$ & $0.639\pm0.157$ & $1.000$ \\ 
& WD 8 & $1.219\pm0.009$ & $0.592\pm0.116$ & $1.000$ & & $1.224\pm0.015$ & $0.628\pm0.146$ & $1.000$ \\ 
& WD 9 & $1.302\pm0.048$ & $0.599\pm0.118$ & $1.000$ & & $1.333\pm0.046$ & $0.845\pm0.220$ & $1.000$ \\   \hline
NGC2168& WD 1 & $4.627\pm0.019$ & $0.954\pm0.016$ & $1.000$ & & $4.625\pm0.019$ & $0.96\pm0.017$ & $1.000$ \\ 
& WD 2 & $5.762\pm0.078$ & $1.068\pm0.015$ & $1.000$ & & $5.764\pm0.079$ & $1.067\pm0.015$ & $1.000$ \\ 
& WD 3 & $5.279\pm0.045$ & $1.020\pm0.012$ & $1.000$ & & $5.277\pm0.047$ & $1.021\pm0.012$ & $1.000$ \\ 
& WD 4 & $4.244\pm0.008$ & $0.916\pm0.022$ & $1.000$ & & $4.243\pm0.008$ & $0.923\pm0.023$ & $1.000$ \\ 
& WD 5 & $4.240\pm0.008$ & $0.915\pm0.022$ & $1.000$ & & $4.239\pm0.008$ & $0.923\pm0.023$ & $1.000$ \\ 
& WD 6 & $5.208\pm0.045$ & $1.013\pm0.012$ & $1.000$ & & $5.203\pm0.046$ & $1.015\pm0.013$ & $1.000$ \\ 
& WD 7 & $6.430\pm0.135$ & $1.134\pm0.020$ & $1.000$ & & $6.450\pm0.138$ & $1.131\pm0.021$ & $1.000$ \\ 
& WD 8 & $4.468\pm0.018$ & $0.938\pm0.018$ & $1.000$ & & $4.465\pm0.020$ & $0.945\pm0.019$ & $1.000$ \\ 
& WD 9 & $4.240\pm0.008$ & $0.915\pm0.022$ & $1.000$ & & $4.239\pm0.008$ & $0.923\pm0.023$ & $1.000$ \\ 
& WD 10 & $5.469\pm0.057$ & $1.039\pm0.013$ & $1.000$ & & $5.467\pm0.059$ & $1.039\pm0.013$ & $1.000$ \\ 
& WD 11 & $5.549\pm0.061$ & $1.047\pm0.014$ & $1.000$ & & $5.546\pm0.062$ & $1.047\pm0.014$ & $1.000$ \\ 
& WD 12 & $4.383\pm0.019$ & $0.930\pm0.019$ & $1.000$ & & $4.378\pm0.020$ & $0.936\pm0.020$ & $1.000$ \\ 
& WD 13 & $4.979\pm0.029$ & $0.990\pm0.013$ & $1.000$ & & $4.976\pm0.030$ & $0.993\pm0.014$ & $1.000$ \\ \hline
NGC2477& WD 1 & $2.348\pm0.033$ & $0.701\pm0.034$ & $1.000$ & & $2.337\pm0.036$ & $0.714\pm0.040$ & $1.000$ \\ 
& WD 2 & $3.114\pm0.031$ & $0.754\pm0.027$ & $1.000$ & & $3.112\pm0.031$ & $0.763\pm0.030$ & $1.000$ \\ 
& WD 3 & $2.954\pm0.028$ & $0.743\pm0.028$ & $0.950$ & & $2.949\pm0.030$ & $0.753\pm0.031$ & $0.964$ \\ 
& WD 4 & $6.951\pm0.654$ & $1.019\pm0.034$ & $1.000$ & & $6.994\pm0.678$ & $1.010\pm0.049$ & $1.000$ \\ 
& WD 5 & $6.471\pm0.835$ & $0.986\pm0.061$ & $1.000$ & & $6.565\pm0.836$ & $0.981\pm0.063$ & $1.000$ \\ 
& WD 6 & $6.147\pm0.981$ & $0.965\pm0.080$ & $1.000$ & & $6.294\pm0.990$ & $0.964\pm0.078$ & $1.000$ \\ 
& WD 7 & $4.402\pm0.735$ & $0.845\pm0.050$ & $1.000$ & & $4.354\pm0.698$ & $0.845\pm0.048$ & $1.000$ \\ \hline
\end{tabular}
\end{table*}

\begin{table}
\centering
\caption{Estimates of the average IFMR parameters from the hierarchical model
and comparisons with other results}
\label{tab:pop}
\vspace{.2cm}
\begin{tabular}{l  ll}
\hline\hline
              & IFMR Intercept & IFMR Slope \\  \hline
  Hierarchical Model & $0.440\pm0.140$ & $0.090\pm0.040$ \\
  \citet{kalirai2008the}    & $0.394\pm0.025$ & $0.109\pm0.007$  \\
  \citet{williams2009} & $0.339\pm0.015$ & $0.129\pm0.004$      \\ \hline
\end{tabular}
\end{table}

Table \ref{tab:pop} presents the estimates of the average IFMR parameters under the
Bayesian hierarchical model and compares them with results from
\citet{kalirai2008the} and \citet{williams2009}.  
In our analysis, we include five clusters: the Hyades, M67, NGC 188, NGC 2168
and NGC 2477. The 68.3\% credible intervals for the IFMR intercept and slope 
are $0.440\pm0.140$ (i.e., $[0.30, 0.58]$) and $0.090\pm0.040$ (i.e., $[0.05, 0.13]$), respectively.
The point estimates of IFMR parameters from \citet{kalirai2008the} and \citet{williams2009}
falls into the credible intervals from our hierarchical model, so we consider that the
average IFMR from our analysis is consistent with results from these studies. 

The point estimate of IFMR intercept from our Bayesian hierarchical analysis is 0.440, greater than
 intercepts from both \citet{kalirai2008the}, 0.394, and \citet{williams2009}, 0.339.
On the other hand, our estimate of the IFMR slope is the shallowest one, 0.090, 
and \citet{williams2009} has the steepest IFMR, at 0.129.

The most obvious characteristic of our result lies in the large error bars, about
6 to 10 times the error bars in other analyses. There are three  
 reasons or this. 1) Only five clusters are included in our analysis. By contrast,
\citet{kalirai2008the} and \citet{williams2009} employed 13 and 11 stellar
clusters, respectively. They had more data to constrain the
IFMR parameters, which leads to the narrow error bars in their studies.
2) In our analysis, clusters have different IFMR parameters and we report
the mean IFMR of these clusters. However, \citet{kalirai2008the} and \citet{williams2009}
assume all clusters have the same IFMR. Their model
is simpler and can be fit with fewer data.
3) \citet{kalirai2008the} and \citet{williams2009} used linear regression
to fit the initial and final masses of WDs in their clusters,
so their estimates of the IFMR parameters are mainly subject to uncertainties of the WDs'
initial and final masses. Their estimates are also indirectly 
affected by the ages and distances of clusters.
 However,
we utilise a Bayesian hierarchical model, which takes account of uncertainties 
of ages, distances, metallicities, etc.

\subsection{Comparison with Spectroscopic Mass Estimates}\label{sec:hydes.com}

In this section, we compare our estimates of initial and final masses of 
Hyades WDs with those determined spectroscopically by
\citet{kalirai2014the}. The meaningful
comparison
is among the final masses of WDs, because that is what \citeauthor{kalirai2014the} are directly
determining with their spectroscopy.

\begin{table*}
\centering
\caption{68.3\% confidence intervals for the
initial and final masses of WDs in Hyades obtained through our hierarchical model and
spectroscopic analysis in \citet{kalirai2014the}}
  \vspace{.2cm}
\label{tab:hyd.mass}
\begin{tabular}{llllllll}
\hline\hline
\multirow{2}{*}{WDs}       & \multirow{2}{*}{Name} & & \multicolumn{2}{c}{Initial Mass} &  & 
\multicolumn{2}{c}{Final Mass} \\ \cline{4-5} \cline{7-8} 
                               & &  & Hierarchical & \citeauthor{kalirai2014the} &   & Hierarchical & 
                               \citeauthor{kalirai2014the} \\ \hline
    0352+096 & HZ 4    &&  $3.779 \pm 0.229$   & $3.59^{+0.21}_{-0.15}$   &&  $0.763 \pm 0.026$   &  $0.80 \pm 0.03$ \\
    0406+169 & LB 227 &&  $3.648 \pm 0.188$   &  $3.49^{+0.13}_{-0.10}$  &&  $0.746 \pm 0.022$   &  $0.85 \pm 0.03$\\
   0421+162 & VR 7    &&   $2.988 \pm 0.079$  & $2.90 \pm 0.02$   &&  $0.657 \pm 0.019$   & $0.70 \pm 0.03$\\
     0425+168 &  VR 16  && $2.813 \pm 0.064$ &  $2.79 \pm 0.01$ &&  $0.632 \pm 0.024$  &   $0.71 \pm 0.03$ \\
    0431+126 &   HZ 7    && $2.942 \pm 0.074$  & $2.84 \pm 0.02$  &&   $0.650 \pm 0.020$  & $0.69 \pm 0.03$ \\
 0438+108  &  HZ 14 &&  $2.801 \pm 0.064$ &  $2.78 \pm 0.01$ && $0.631 \pm 0.024$  & $0.73 \pm 0.03$  \\
      \hline     
  %   \multicolumn{8}{l}{$^{a}$ is short for \cite{kalirai2014the}.} 
\end{tabular}
\end{table*}

\begin{figure*}
\centering
\includegraphics[width=0.9\textwidth]{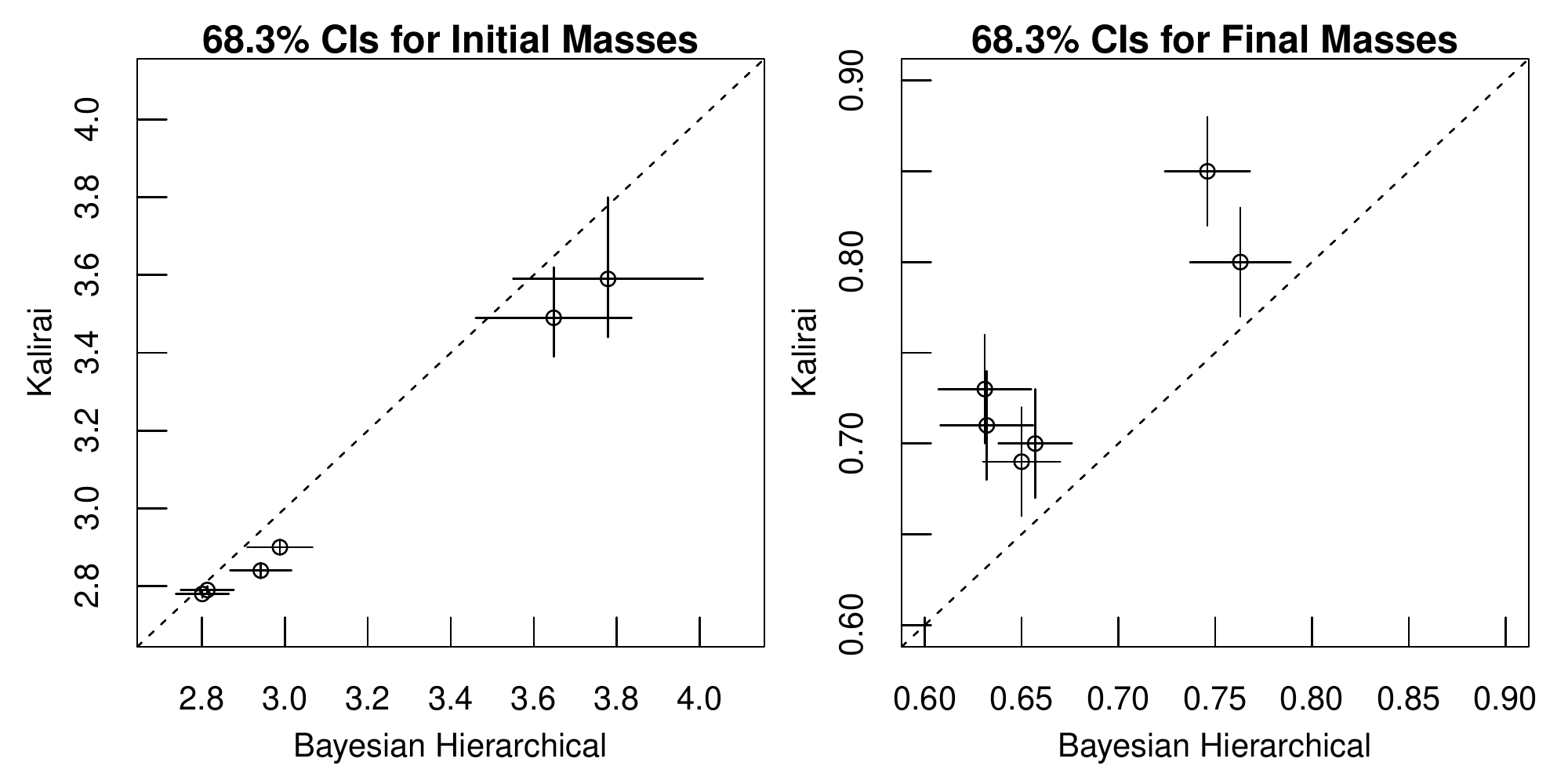}
\caption{The 68.3\% confidence intervals for initial and final masses of the Hyades WDs from our
Bayesian hierarchical modelling and \citet{kalirai2014the}. The left panel compares the
 initial masses of the two approaches and 
the right panel compares the final masses. Horizontal and vertical error bars
represents the 68.3\% CIs 
from our hierarchical analysis and \citet{kalirai2014the}. 
\label{fig:hyd.mass}}
\end{figure*}

Table \ref{tab:hyd.mass} and Fig. \ref{fig:hyd.mass} present the 68.3\% confidence intervals (CIs) for 
these initial and final WD masses.
The left panel presents the initial masses and 
the right panel presents the final masses. In each of these plots, bars parallel to 
the x-axis and y-axis are 68.3\% CIs 
from our hierarchical analysis and \citet{kalirai2014the}, respectively. 
The BASE-9 results in Fig. \ref{fig:hyd.mass} are 
$\sim0.05$ to $0.10 M_{\odot}$ lower than the spectroscopic
results from \citeauthor{kalirai2014the}.  Interestingly, both are ultimately based on the Montreal
white dwarf group's models, though the spectroscopic technique relies on the
Balmer line profiles whereas the photometric technique relies on the overall
SED of the WD.  While these differences are small, at least in this case they
appear systematic, and may indicate subtle inconsistencies between the model
colours and line profiles.  Alternatively, the photometric technique relies
on the cluster distance, which may be slightly in error and will be improved
upon with Gaia results \citep{babusiaux2018gaia,lindegren2018gaia}.

\subsection{Comparison with Gaia Estimates of Distance Moduli}\label{sec:gaia.com}

As this paper was completed, the Gaia Collaboration \citep{babusiaux2018gaia,lindegren2018gaia} 
analysed all clusters in this paper with the exception of NGC 2477,
so we revisit the distributions of distance moduli in our analysis and
 the Gaia results.

\citet{babusiaux2018gaia} employed the Hertzsprung-Russell diagrams (HRD) to study
stars with data from Gaia Data Release 2 (Gaia DR2) and presented several illustrative
examples. \citet{lindegren2018gaia} updates the results in \citet{babusiaux2018gaia} 
by showing a global parallax zero point of about
$-0.029$ milli-arcsec (mas). Here we collect the published parallaxes and their standard errors
of these four clusters  from Table A.3 and A.4 in \citet{babusiaux2018gaia},
take $0.01$ mas as the minimum error based on uncertainty in the Gaia parallax zero point
 and compute their distance moduli and 
standard errors as shown in Table \ref{tab:gaia.coll}.
We follow the correction from \citet{lindegren2018gaia}, add $0.029$ mas to the published parallaxes
in \citet{babusiaux2018gaia}, take $0.01$ mas as the minimum error of parallaxes and
obtain the corrected estimates of distance moduli presented in Table \ref{tab:gaia.coll}.
 
Table \ref{tab:gaia.coll} summarises the prior and posterior distributions of distance
moduli for four clusters (Hyades, M67, NGC 188 and NGC 2168) in our analysis and the results
from the Gaia Data Release 2 (Gaia DR2). 
\begin{table*}
\centering
\caption{The prior and posterior distributions of \textbf{distance moduli} of four clusters (Hyades, M67, NGC 188
and NGC 2168) in our analysis
and results based on Gaia DR2 \citep{babusiaux2018gaia,lindegren2018gaia}. The resulting
$z_{1}$-scores and $z_{2}$-scores are statistical differences between prior distributions 
and Gaia estimates from \citet{babusiaux2018gaia} and \citet{lindegren2018gaia}, respectively.}
  \vspace{.2cm}
\label{tab:gaia.coll}
\begin{tabular}{lllllll}
\hline\hline
{Cluster} &  {Prior} &  {Posterior} & \citeauthor{babusiaux2018gaia} & \citeauthor{lindegren2018gaia} & $z_{1}-$score & $z_{2}-$score\\ \hline 
  Hyades    &  $3.40 \pm 0.03$   & $3.40 \pm 0.03$   & $3.384 \pm 0.007$ &$3.381 \pm 0.007$ & $0.52$ & $0.62$\\
   M67    &   $9.62 \pm 0.091$  & $9.85 \pm 0.01$   & $9.73 \pm 0.019$ & $9.675 \pm 0.019$&$1.18$ & $0.59$ \\
   NGC 188  & $11.24 \pm 0.1$ &  $11.51 \pm 0.01$ &  $11.482 \pm 0.043$ &$11.361 \pm 0.041$& $2.22$ & $1.12$ \\
   NGC 2168   & $10.3 \pm 0.1$  & $10.29 \pm 0.01$  &   $9.756 \pm 0.019$ &$9.70 \pm 0.019$& $5.34$ & $5.89$\\
      \hline     
\end{tabular}
\end{table*}
The $z_{1}, z_{2}$-scores are statistical measures of the consistency between the prior distributions
of distance moduli in our analysis and results based on \citet{babusiaux2018gaia} and
\citet{lindegren2018gaia}, respectively.
If the z-score is less than or equal to 1.96, it means that the prior distribution we used is consistent to
the Gaia result under the significance level 0.05. Otherwise, the two are significantly different.
Therefore, the prior distributions of distance moduli for Hyades and M67 are consistent with the Gaia
results from \citet{babusiaux2018gaia} and \citet{lindegren2018gaia}. 
For NGC 188, the prior distribution in our analysis is significantly different from the estimate from
\citet{babusiaux2018gaia}, but it is consistent with the corrected distance modulus estimate in
\citet{lindegren2018gaia}.
For NGC 2168, its distance modulus prior distribution is significantly different from estimates
from both \citet{babusiaux2018gaia} and \citet{lindegren2018gaia}.

The posterior distributions in our analysis are also shown in Table \ref{tab:gaia.coll}.
For NGC 188, we used $N(11.24, 0.1^2)$ as its distance modulus 
prior distribution and it yielded a distance modulus posterior distribution $N(11.51, 0.01^2)$,
differing significantly from the prior, which means that the posterior distribution is dominated
by the photometric data rather than the prior. Therefore, the prior distribution does not matter much for
NGC 188. Interestingly, our distance modulus posterior distribution for NGC 188 is close
to the result from \citet{babusiaux2018gaia}. So we believe the joint posterior distribution
for NGC 188 is  unlikely to change even if we used its Gaia distance modulus estimate as 
the prior.  

In summary, three (the Hyades, M67 and NGC 188) of the four clusters would 
most likely be unchanged with the Gaia distance priors.  The other one (NGC 2168) 
most likely would be changed, but without a full re-analysis,
which will be accomplished in the future with more clusters, it is hard to know how 
this would affect the hierarchical IFMR results.
We will very likely redo the hierarchical analysis with the Gaia distance moduli as prior
distributions when 10 or more clusters are available to us in the near future.

\section{Sensitivity Analysis}\label{sec:sen}

In this section, we present the sensitivity analysis in the hierarchical
analysis of IFMR parameters.

\subsection{Sensitivity to Prior Distribution} \label{sec:ifmr.prior.sensi}

Here we investigate whether the hierarchical analysis in Eq. \ref{eq:hier}
is sensitive to the prior distribution on $\bm{\Gamma}$.
We use
the marginally non-informative prior distribution proposed by \citet{huang2013simple}, i.e.,
\begin{align*}
&\bm{\Gamma}\mid\lambda_{1}, \lambda_{2}\sim\mbox{Inverse Wishart}
\Bigg(2\nu
\begin{pmatrix}
1/\lambda_{1} & 0\\
0 & 1/\lambda_{2}
\end{pmatrix},~~\nu +1\Bigg), \\
&\lambda_{1}, \lambda_{2}\sim\mbox{Inverse Gamma}(1/2,  1/\Lambda),
\end{align*}

$\lambda_{1}$ and $\lambda_{2}$ are hyper-parameters, and they
independently follow the same inverse gamma distribution with its first
parameter fixed at $1/2$ and second parameter $1/\Lambda$ a small positive number,
i.e., large positive $\Lambda$.
\citet{huang2013simple} showed that $\nu = 2$ leads to a marginal uniform
distribution for correlation $\rho$ and arbitrarily large positive $\Lambda$ leads to 
arbitrarily weakly informative prior distributions for $\sigma_{1}$ and $\sigma_{2}$.
Because $\nu = 2$ is necessary to have a marginally non-informative
prior distribution on $\rho$, so in this hierarchical analysis, we take $\nu=2$.
As for $\Lambda$, we choose four large values: $10^3, 10^4, 10^5, 10^6$ and
fit the hierarchical model with these values, then compare the MCMC draws of
IFMR parameters of the five included clusters.

\begin{figure*}
\centering
\includegraphics[width=.6\textwidth]{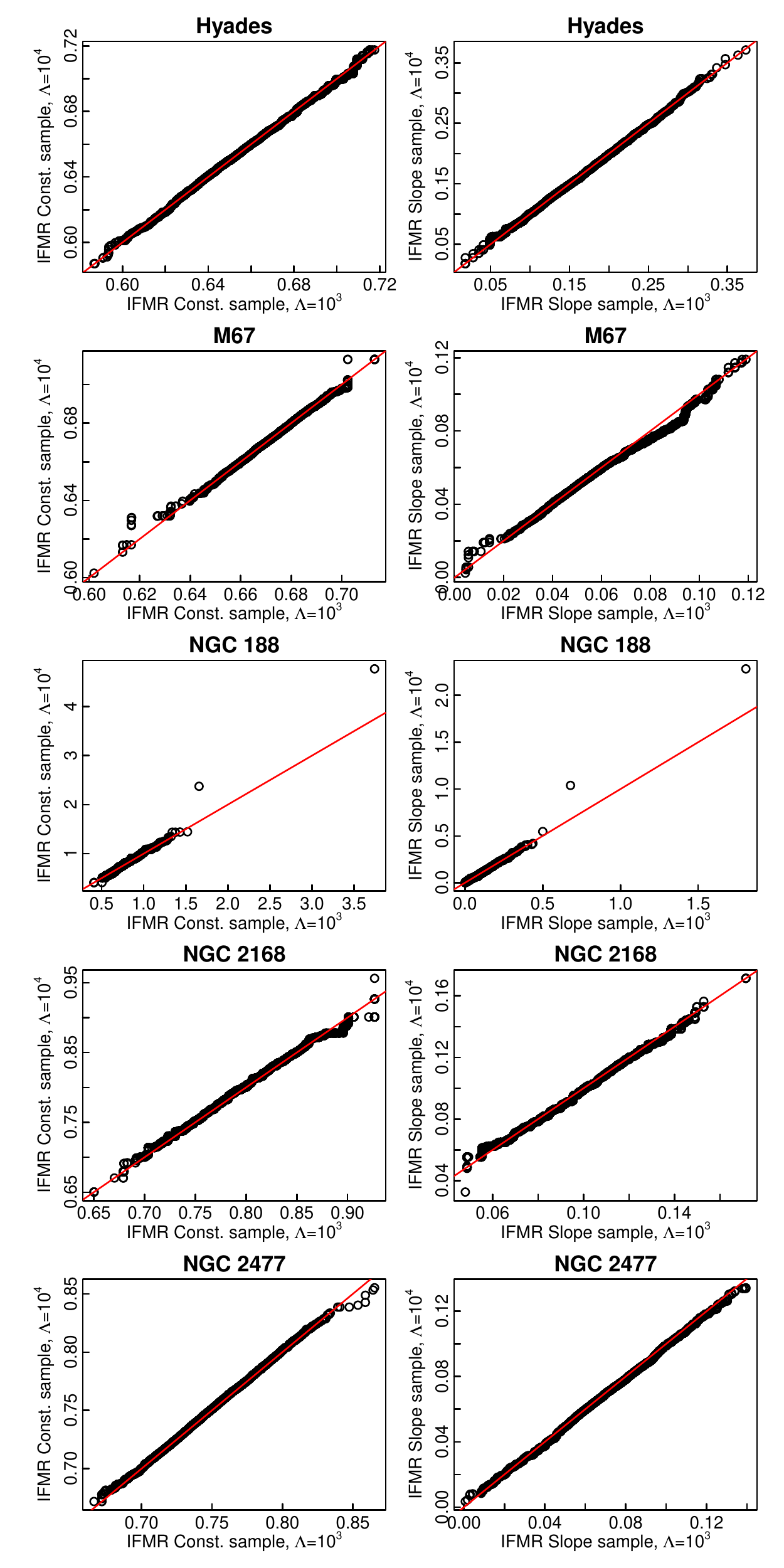}
\caption{ Quantile-quantile plots of MCMC samples of IFMR parameters
of five clusters. The x-axis and y-axis are quantiles corresponding to fits
with $\Lambda = 10^3$ and $\Lambda = 10^4$, respectively.
The $45^\circ$ red line represents
that quantiles from samples in x and y axes are equal. 
  \label{fig:ifmr.prior.sen1}}
\end{figure*}

\begin{figure*}
\centering
\includegraphics[width=.6\textwidth]{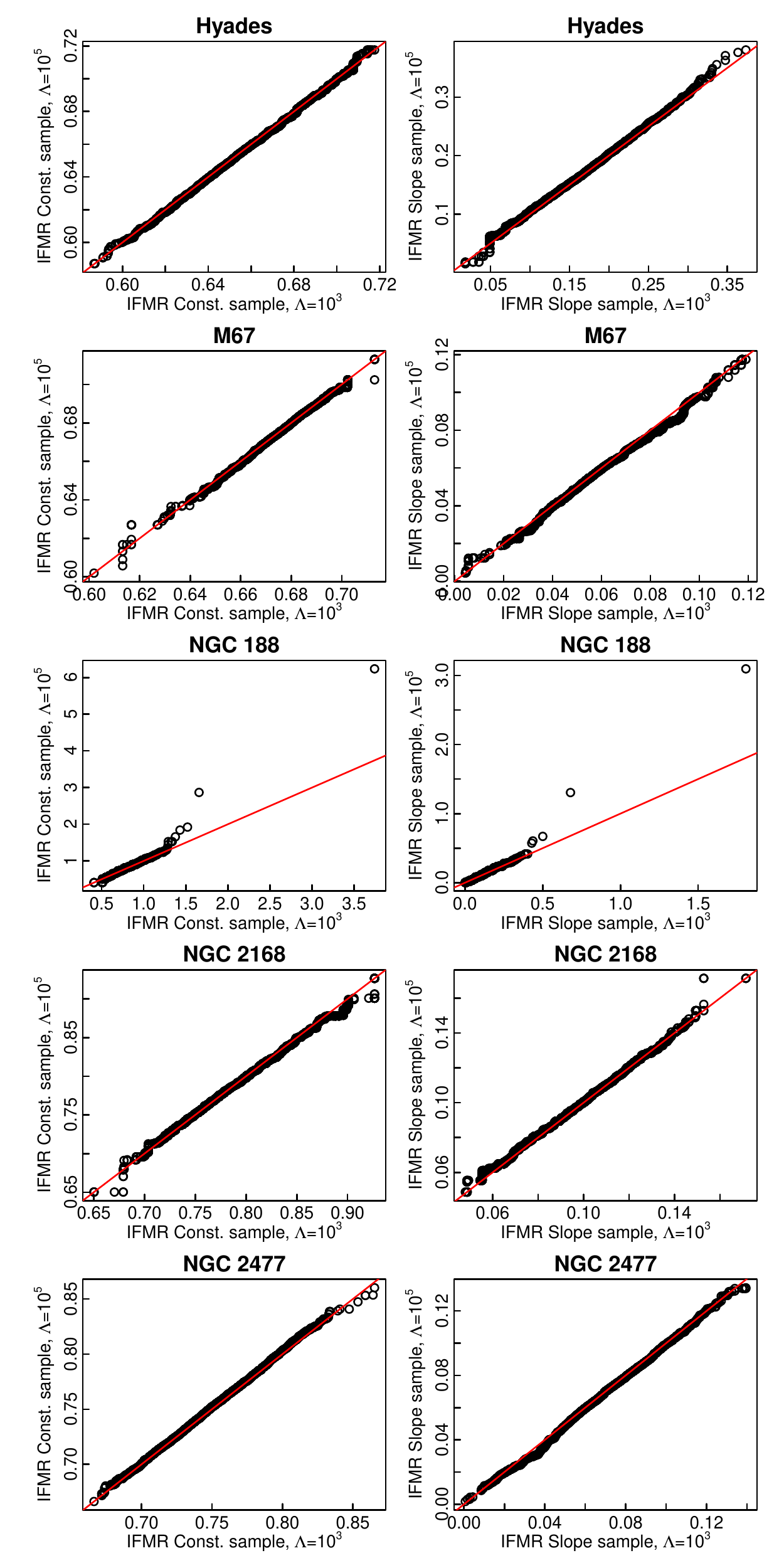}
\caption{ Quantile-quantile plots of MCMC samples of IFMR parameters
of five clusters. The x-axis and y-axis are quantiles corresponding to fits
with $\Lambda = 10^3$ and $\Lambda = 10^5$, respectively.
The $45^\circ$ red line represents
that quantiles from samples in x and y axes are equal. 
  \label{fig:ifmr.prior.sen2}}
\end{figure*}

\begin{figure*}
\centering
\includegraphics[width=.6\textwidth]{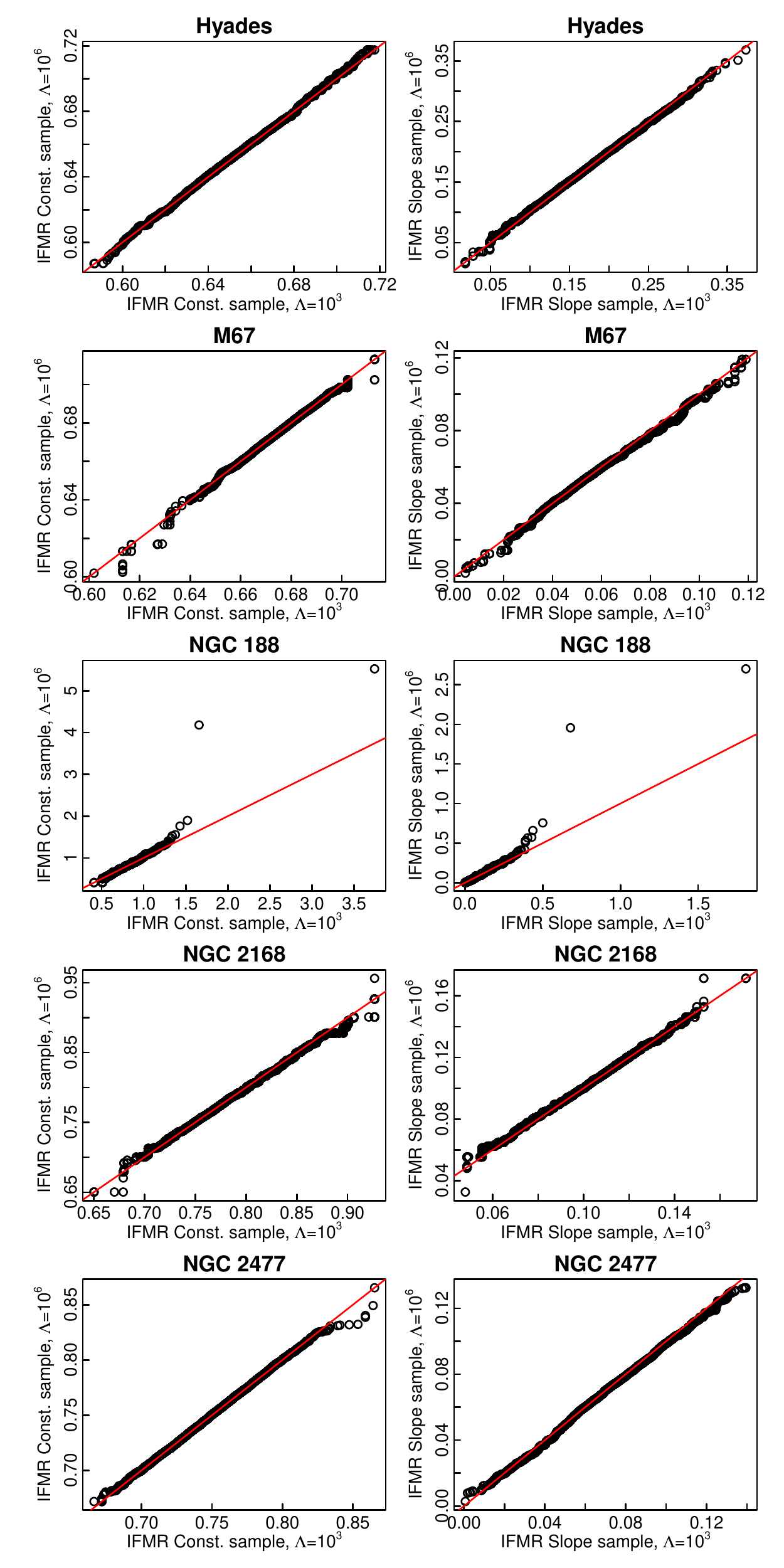}
\caption{ Quantile-quantile plots of MCMC samples of IFMR parameters
of five clusters. The x-axis and y-axis are quantiles corresponding to fits
with $\Lambda = 10^3$ and $\Lambda = 10^6$, respectively.
The $45^\circ$ red line represents
that quantiles from samples in x and y axes are equal. 
  \label{fig:ifmr.prior.sen3}}
\end{figure*}

Fig.s \ref{fig:ifmr.prior.sen1}--\ref{fig:ifmr.prior.sen3} present the QQ plots of
IFMR parameters from the hierarchical fits when $\Lambda$ takes different
values. All points in these QQ plots lie close to the $45^\circ$ red line, meaning that
MCMC draws of IFMR parameters are essentially the same. \label{sen:deg.sym}
Though there are some small deviations from the red line, they are mainly
caused by Monte Carlo errors. So
we conclude that the hierarchical result is not sensitive to the choice
of value of $\Lambda$ provided that $\Lambda \geq 10^3$.
  
\subsection{Sensitivity to Membership of WDs in M67} \label{sec:ifmr.wd.sensi}

Table \ref{tab:mass2} presents the hierarchical and case-by-case
estimates of initial and final masses and membership probabilities for 35 WDs in M67, among which
nine WDs have posterior membership probabilities less than $0.5$ and are classified
as non-members or field stars. The other 26 WDs are inferred as members of
M67.
In this section we investigate whether the membership of these nine WDs affects 
the case-by-case posterior distribution of cluster M67.
 
In Section \ref{sec:data}, when performing the case-by-case analysis on the cluster M67
with BASE-9, we set the prior membership probabilities of all WDs based on other research
\citep[e.g.,][]{bellini2010absolute,bellini2010end,williams2013time,barnes2016rotation}.
For clarity, we call this analysis the original fit.
Then we fit M67 under two other circumstances: 1.) Case I:
setting all WDs to have a 100\% prior probability of being a member in M67,
and 2.) Case II: assigning nine non-members as determined by
 the original fit (Table \ref{tab:mass2})
to have prior membership probabilities equal to 0 and the apparent cluster members
 to have prior probabilities equal to 1.
In other words, the first case forces all WDs in M67 to be cluster members whereas
the second case removes nine apparent non-members and assumes that
 the other 26 as definitive cluster members.   

Table \ref{tab:ifmr.wd.sen} presents the point estimates
of M67 parameters under three settings. 
The last two columns present the IFMR constant and slope, respectively.
The case-by-case estimates of the
IFMR slope vary under different settings.
In Case I, when all WDs are forced to be M67 members, the IFMR slope is the steepest,
while in the original fit, the IFMR is the shallowest.
The estimates of other parameters are similar under the settings except the age.
The estimate of age from the original fit is consistent to that from the Case II,
while the age estimate from Case I is younger than the others.
The age estimates from the original fit and Case I
are close to the results obtained through other approaches
 \citep{bellini2010end,williams2013time}.
 
\begin{table*}
\centering
\caption{Estimated cluster parameters for M67 under three cases.}
\label{tab:ifmr.wd.sen}
\vspace{.2cm}
\begin{tabular}{lllllll}
\hline\hline
   Settings     & Age (Gyr) & $m - M_{V}$ & [Fe/H]$^{1}$ & Abs.$^{1}$ & IFMR Constant & IFMR Slope \\  \hline
  Original & $3.90\pm0.02$ & $9.85\pm0.01$ & $-0.03$ & $0.14$ & $0.68\pm0.01$ & $0.05\pm0.01$ \\
  Case I   & $3.07\pm0.03$ & $9.86\pm0.02$ & $-0.01$ & $0.15$ & $0.67\pm0.01$ & $0.10\pm0.01$ \\
  Case II & $3.89\pm0.02$ & $9.86\pm0.01$ & $-0.04$ & $0.16$ & $0.68\pm0.01$ & $0.07\pm0.01$ \\
    \hline
\multicolumn{7}{l}{$^{1}$: The standard errors for metallicity and absorption under these fits}\\
\multicolumn{7}{l}{are all 0.01.}    
\end{tabular}
\end{table*}

\begin{figure*}
\centering
\includegraphics[width=\textwidth]{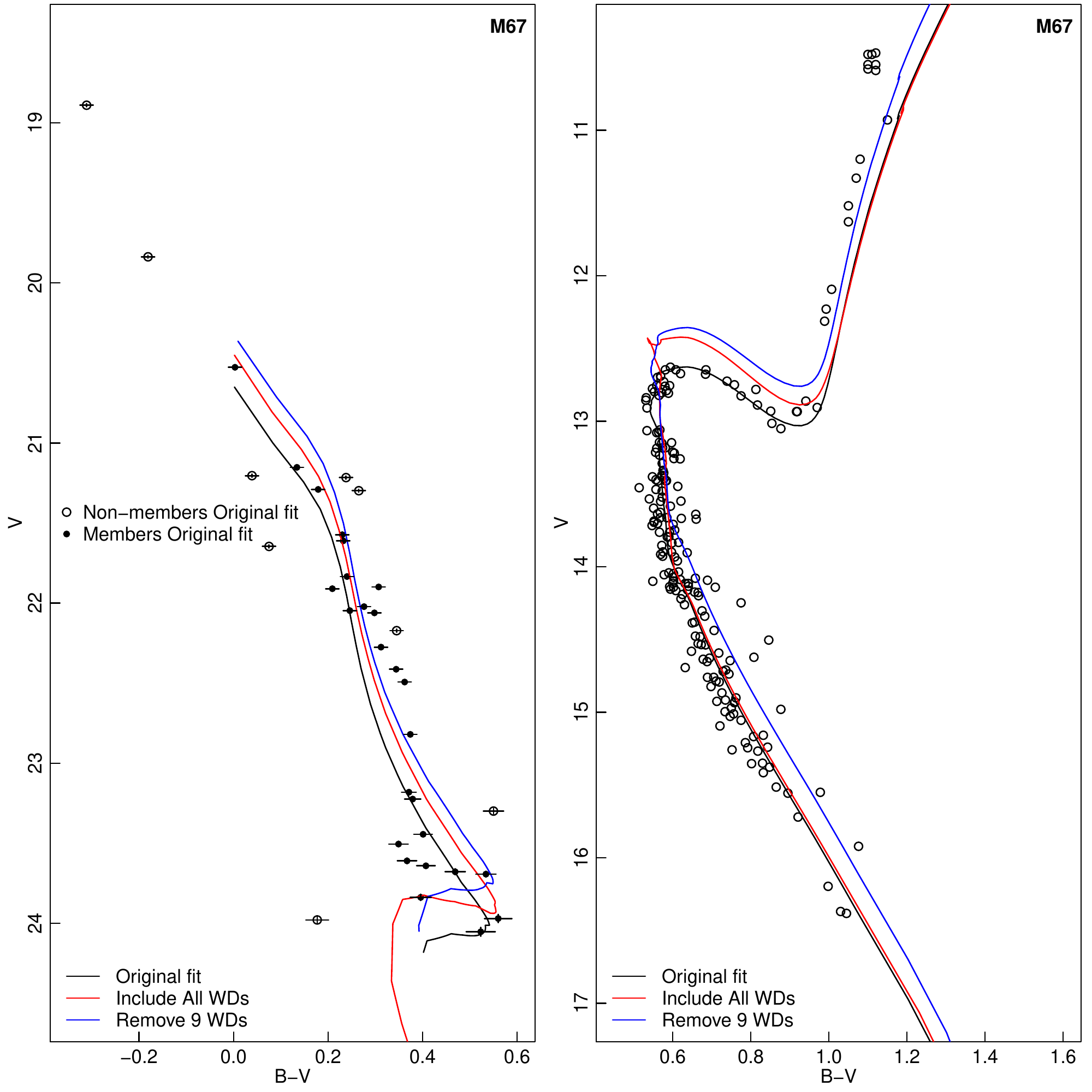}
\caption{Colour-magnitude diagram (CMD) plots of three fits.
The black lines are from the original fit,
red lines from the fit that includes all WDs in M67,
blue lines from the fit that removes nine non-members. 
In the left panel, open circles represent non-member WDs 
from the original fit and solid points are members. 
Observed error bars are included for the WDs.
  \label{fig:ifmr.wd.sen}}
\end{figure*}

Fig. \ref{fig:ifmr.wd.sen} presents the CMD plots for these three fits.
The black lines are from the fitted model under the original fit, and
the red and blue lines are from models under Case I and II, respectively.
From this plot, both fits from Case I (red line) and II (blue line)
miss the main sequence turnoff stars, sub-giant branch and base
of the red giant branch. By contrast, the model under the original fit (black line)
matches both parts of the cluster well.
We conclude that the original fit, where BASE-9 was able to assign its own cluster 
membership probabilities, is more reliable than the other fits.
In summary, the membership of these WDs affect the posterior 
distribution of cluster parameters, yet our further analysis supports the original fit because it 
best matches the photometric data among those three fits.

\subsection{Sensitivity to WD-WD Binaries} \label{sec:wd.binary}

Our BASE-9 model does not include WD-WD binaries.
While it will likely be preferable to do so eventually, current studies of cluster WDs are
inadequate to determine the fraction of double degenerates in clusters and
even further from determining which cluster WDs are unresolved binaries. The
possible exceptions to this are the Hyades WDs, which are nearby, relatively
bright, and well-studied.  Among the 7 Hyades in our study, it is likely
that all are single WDs.  Theoretical studies \citep[e.g.,][]{hurley2005a}
indicate that the number of unresolved WD-WD binaries is probably
$<10$\% of a cluster's WD population.
Thus $\sim3$ of the M67 WDs and $\sim1$ each for NGC 188, NGC 2168, and NGC
2477 may be unresolved double degenerates.  The WD regions of the CMDs for
all of these clusters are consistent with this possibility.  Fortunately,
BASE-9 is robust against a small fraction of WDs having a large effect on
the IFMR fit because (a) the double degenerate fraction is likely to be
small and (b) objects that fall overly far from the best fit isochrones are
fit as non-members and therefore do not contribute to the cluster solution
(age, WD mass, IFMR parameters).

\section{Conclusions and Discussions}\label{sec:con}

We proposed a Bayesian hierarchical model for the IFMR parameters that
simultaneously analyses data from multiple clusters in a single overall model and 
produces more precise estimates of the IFMR parameters.
Also, we develop an efficient two-stage algorithm that takes advantage of existing
software for cluster-specific analysis to
obtain the fit under the hierarchical model.
We combine data from five open clusters in the Bayesian hierarchical model and find that 
it can correct an error in the estimates of IFMR parameters for the
 cluster NGC 188 and produce reasonable estimates of IFMR for NGC 188.
Based on our hierarchical analysis, we estimate
 the linear IFMR averaged across clusters to be
$$M_{\rm final}=(0.09\pm0.04)M_{\rm initial}+(0.44\pm0.14)M_{\sun},$$
with $0.8M_{\sun}\leq~M_{\rm initial}\leq8.0 M_{\sun}$.

This paper focus on the use of statistical techniques to the IFMR project, and
the detailed results are preliminary. 
In particular, the astronomical results in this paper are not definitive and
they depend upon the models inside the black-box code (in our case, the BASE-9) and
other assumptions.
Specifically, we assumed that the IFMR parameters from different clusters 
follow a bivariate normal distribution.
However, this assumption might be too idealised. Bayesian hierarchical models
always require a population distribution on all objects and the
 shape of the distribution of IFMR parameters across all clusters is not available.
We therefore use the bivariate normal distribution as a starting point.
If a case can be made for a different distribution, 
the statistical algorithm developed in this chapter will work as long as the
case-by-case results are valid.
Studies have indicated that
 metallicity may affect the IFMR
\citep[See, e.g.,][]{kalirai2005the,catalan2008the,meng2008initial,zhao2012the}.
We showed how the Bayesian hierarchical model in Eq. \eqref{eq:hier} can be readily extended to 
investigate the effect of metallicity.
In addition, even though our research is based on the BASE-9 package, our statistical techniques
 can be employed with other black-box packages as long
as they produce MCMC samples in their case-by-case fits.
Different underlying stellar evolution models will affect the results of case-by-case
fits, hence they will most likely impact the resulting hierarchical fit.
Our statistical approach and computational algorithm are independent of these inputs and 
can be broadly applied.

 \section*{Acknowledgements}
We thank the Imperial College High Performance Computing support team for their kind help,
 which greatly 
accelerated the simulation study in this project. Shijing Si thanks the Department of Mathematics in
Imperial College London for a Roth studentship, which supported his research. 
David van Dyk acknowledges support 
from a Marie-Skodowska-Curie RISE Grant 
(H2020-MSCA-RISE-2015-691164) provided by the European Commission. 
Ted von Hippel acknowledges support from the National Science Foundation under Award AST-1715718.
The authors gratefully thank the Referee and editors for the constructive comments and 
recommendations which definitely helped to improve the readability and quality of the paper. 
All the comments are addressed accordingly and have been incorporated to the revised manuscript.

\bibliographystyle{mnras}
\bibliography{all}

%%%%%%%%%%%%%%%%%%%%%%%%%%%%%%%%%%%%%%%%%%%%%%%%%%

%%%%%%%%%%%%%%%%% APPENDICES %%%%%%%%%%%%%%%%%%%%%

\appendix

\section{Computational Algorithm}\label{app:alg}

The joint posterior distribution in Eq.~\ref{eq:jp} is high-dimensional, with $2N\times{K}+6K+7$ parameters.
In this appendix, we 
show how to take advantage of the cluster-specific fittings in BASE-9 via a 
two-stage (TS) algorithm to fit the hierarchical model.

To simplify the description of the TS algorithm,
we introduce $\bm{\Omega}_{k}=(\bm{M}_{k}, \bm{R}_{k}, \bm{Z}_{k}, \bm{\Theta}_{k})$
for $k= 1, 2, \ldots, K$. We use $\bm{\Psi}={\rm diag}(1/\lambda_{1}, 1/\lambda_{2})$.

\begin{description}
\item[\textbf{Step 0a:}] For each star cluster run BASE-9 to obtain a Monte Carlo sample of
${p}(\bm{M}_{k}, \bm{R}_{k}, \bm{\alpha}_{k}, \bm{Z}_{k}, \bm{\Theta}_{k}|\bm{X}_{k}, \bm{\Sigma}_{k})$
via the 
 cluster-specific analysis. Thin each chain to obtain an essentially independent Monte Carlo sample
 and label it $\{\bm{\Omega}_{1}^{(t)}, \bm{\alpha}^{(t)}_{1}, 
 \cdots, \bm{\Omega}_{K}^{(t)}, \bm{\alpha}^{(t)}_{K}, t=1, 2, \cdots, t_{MC}\}$. 
 \item[\textbf{Step 0b:}] In the following, we denote the TS samples with the tilde notation.
 Simulate $K$ random integers between $1$ and $t_{MC}$,
 denote them $r_{1}^{\ast}, \ldots, r_{K}^{\ast}$, and
 initialise the parameters $\tilde{\bm{\alpha}}_{k}^{(1)}=\bm{\alpha}_{k}^{(r_{k}^{\ast})}$ and 
$\tilde{\bm{\Omega}}_{k}^{(1)}=\bm{\Omega}_{k}^{(r_{k}^{\ast})}$ for $k= 1, \ldots, K$.\\
\item[\textbf{Step 0c:}] Given $\tilde{\bm{\Omega}}_{1}^{(1)}, \tilde{\bm{\alpha}}_{1}^{(1)}, \ldots,
\tilde{\bm{\Omega}}_{K}^{(1)}, \tilde{\bm{\alpha}}_{K}^{(1)}$, simulate $\tilde{\bm{\gamma}}^{(1)}$
 and $\tilde{\bm{\Gamma}}^{(1)}$ via the partially collapsed Gibbs (PCG) sampler \citep{van2008partially}, 
 \begin{align*}
 &\tilde{\bm{\Gamma}}^{(1)}\sim{\rm Inverse~Wishart}\bigg(K+\nu, \sum_{k=1}^{K}
 (\tilde{\bm{\alpha}}_{k}^{(1)}-\tilde{\bm{\alpha}}^{(1)}) (\tilde{\bm{\alpha}}_{k}^{(1)}-\tilde{\bm{\alpha}}^{(1)})^{\top}\bigg),\\
 &\tilde{\bm{\gamma}}^{(1)}\sim{\rm N}_{2}\bigg(\tilde{\bm{\alpha}}^{(1)}, \tilde{\bm{\Gamma}}^{(1)}/K\bigg),
 \end{align*}
 with $\tilde{\bm{\alpha}}^{(1)}=\frac{1}{K}\sum_{k=1}^{K}\tilde{\bm{\alpha}}_{k}^{(1)}$.
 \item[\textbf{Step 0d:}] Given $\tilde{\bm{\Gamma}}^{(1)}$, we simulate $\lambda_{1}, \lambda_{2}$ from their  
  conditional posterior distributions,
$$ \tilde{\lambda}_{\ell}^{(1)}\sim{\rm Inverse~Gamma}\bigg(1+\nu/2, \frac{1}{\Lambda_{\ell}}+\nu(\tilde{\bm{\Gamma}}^{(1)})^{-1}_{\ell\ell}\bigg)$$
 for $\ell= 1, 2$.\\
For $s=1, \cdots$, run Step 1 and Step 2 iteratively.
 %   \begin{description}
      \item[\textbf{Step 1:}] Randomly generate $K$ integers between $1$ and $t_{MC}$, 
      and denote them $r_{1},\ldots, r_{K}$.
For each $k= 1, \ldots, K$, set  $\bm{\alpha}_{k}^{\ast}=\bm{\alpha}_{k}^{(r_{k})}$ and 
$\bm{\Omega}_{k}^{\ast}=\bm{\Omega}_{k}^{(r_{k})}$ as the new proposal and
set $\tilde{\bm{\alpha}}_{k}^{(s+1)}=\bm{\alpha}_{k}^{\ast}, \tilde{\bm{\Omega}}^{(s+1)}=\bm{\Omega}_{k}^{\ast}$ with probability
$\min\bigg\{1, \frac{{p}(\bm{\alpha}_{i}^{\ast}|\tilde{\bm{\gamma}}^{(s)}, \tilde{\bm{\Gamma}}^{(s)})}
{{p}(\bm{\alpha}_{i}^{(s)}|\tilde{\bm{\gamma}}^{(s)}, \tilde{\bm{\Gamma}}^{(s)})}\bigg\}$.
Otherwise, set $\tilde{\bm{\alpha}}_{k}^{(s+1)}=\tilde{\bm{\alpha}}_{k}^{(s)}, \tilde{\bm{\Omega}}^{(s+1)}=\tilde{\bm{\Omega}}^{(s)}$.
         
         \item[\textbf{Step 2:}] Given $\tilde{\lambda}_{\ell}^{(s)}, \ell=1, 2$ and $\tilde{\bm{\alpha}}_{k}^{(s+1)},~
      \text{and}~\tilde{\bm{\Omega}}_{k}^{(s+1)}, k=1, \ldots, K$,
       update $\tilde{\bm{\gamma}}, \tilde{\bm{\Gamma}}$ via
      \begin{align*}
 &\tilde{\bm{\Gamma}}^{(s+1)}\sim{\rm Inverse ~Wishart}\bigg(K+\nu, \\
 &~\sum_{k=1}^{K}
 \Big(\tilde{\bm{\alpha}}_{k}^{(s+1)}-\tilde{\bm{\alpha}}^{(s+1)}\Big)\Big(\tilde{\bm{\alpha}}_{k}^{(s+1)}-\tilde{\bm{\alpha}}^{(s+1)}\Big)^{\top}+2\nu{\rm diag}\Big(1/\tilde{\lambda}^{(s)}_{1}, 1/\tilde{\lambda}^{(s)}_{2}\Big)\bigg),\\
 &\tilde{\bm{\gamma}}^{(s+1)}\sim{\rm N}_{2}\Big(\tilde{\bm{\alpha}}^{(s+1)}, \tilde{\bm{\Gamma}}^{(s+1)}/K\Big),
 \end{align*}
 with $\tilde{\bm{\alpha}}^{(s+1)}=\frac{1}{K}\sum_{k=1}^{K}\tilde{\bm{\alpha}}_{k}^{(s+1)}$. \\
 Given $\tilde{\bm{\Gamma}}^{(s+1)}$, simulate $\tilde{\lambda}_{\ell}, \ell=1,2$ via
 $$ \tilde{\lambda}_{\ell}^{(s+1)}\sim{\rm Inverse~Gamma}\bigg(1+\nu/2, \frac{1}{\Lambda_{\ell}}+\nu(\tilde{\bm{\Gamma}}^{(s+1)})^{-1}_{\ell\ell}\bigg).$$
\end{description}

\section{Photometry Data for WDs in Five Star Clusters}\label{app:phot}

Here are the photometry data for the WDs in these five star
clusters in Section \ref{sec:data}.

 \begin{table*}
 \caption{Photometry for the WDs in the Hyades}
\label{tab:phot}
 \begin{tabular}{ll lll}
 \hline\hline
Cluster & WD & $U\pm\sigma_{U}$ & $B\pm\sigma_{B}$ & $V\pm\sigma_{V}$\\ \hline
Hyades& HZ14 & $9.155\pm0.032$ & $10.195\pm0.030$ & $10.351\pm0.029$ \\ 
& VR16 & $9.298\pm0.114$ & $10.274\pm0.114$ & $10.373\pm0.113$ \\ 
& HZ7 & $9.969\pm0.025$ & $10.859\pm0.022$ & $10.903\pm0.016$ \\ 
& VR7 & $10.129\pm0.049$ & $10.977\pm0.041$ & $11.004\pm0.035$  \\ 
& HZ4 & $11.136\pm0.054$ & $11.811\pm0.054$ & $11.725\pm0.054$ \\ 
& LB227 & $11.034\pm0.015$ & $11.752\pm0.015$ & $11.697\pm0.014$ \\ \hline
\multicolumn{5}{l}{References: \citet{degennaro2009inverting,stein2013}}
\end{tabular}
\end{table*}

 \begin{table*}
 \caption{Photometry for the WDs in M67.}
\label{tab:phot1}
 \begin{tabular}{ll ll}
 \hline\hline
Cluster & WD  & $B\pm\sigma_{B}$ & $V\pm\sigma_{V}$ \\ \hline
M67 & WD 1 &  $23.854\pm0.010$ & $23.505\pm0.018$ \\ 
& WD 2  & $22.293\pm0.010$ & $22.047\pm0.010$  \\ 
& WD 3  & $22.119\pm0.010$ & $21.910\pm0.010$  \\ 
& WD 4  & $21.721\pm0.010$ & $21.646\pm0.010$  \\ 
& WD 5  & $19.656\pm0.010$ & $19.837\pm0.010$ \\ 
& WD 6  & $22.074\pm0.010$ & $21.834\pm0.010$ \\ 
& WD 7  & $21.843\pm0.010$ & $21.611\pm0.010$  \\ 
& WD 8  & $21.469\pm0.010$ & $21.290\pm0.010$  \\ 
& WD 9  & $21.286\pm0.010$ & $21.152\pm0.010$  \\ 
& WD 10  & $23.603\pm0.010$ & $23.224\pm0.014$ \\ 
& WD 11  & $22.206\pm0.010$ & $21.899\pm0.010$ \\ 
& WD 12  & $21.454\pm0.010$ & $21.216\pm0.010$  \\ 
& WD 13  & $20.530\pm0.010$ & $20.527\pm0.010$  \\ 
& WD 14  & $23.194\pm0.010$ & $22.820\pm0.010$  \\ 
& WD 15  & $22.587\pm0.010$ & $22.275\pm0.010$  \\ 
& WD 16  & $24.227\pm0.010$ & $23.693\pm0.019$  \\ 
& WD 17  & $22.359\pm0.010$ & $22.061\pm0.010$  \\ 
& WD 18  & $22.855\pm0.010$ & $22.493\pm0.010$  \\ 
& WD 19  & $23.552\pm0.010$ & $23.181\pm0.012$  \\ 
& WD 20 & $21.243\pm0.010$ & $21.204\pm0.010$  \\ 
& WD 21 & $18.579\pm0.010$ & $18.890\pm0.010$  \\ 
& WD 22  & $24.530\pm0.013$ & $23.970\pm0.026$  \\ 
& WD 23 & $24.576\pm0.013$ & $24.053\pm0.028$  \\ 
& WD 24  & $21.563\pm0.010$ & $21.298\pm0.010$  \\ 
& WD 25  & $23.977\pm0.010$ & $23.610\pm0.018$  \\ 
& WD 26 & $24.147\pm0.010$ & $23.678\pm0.019$  \\ 
& WD 27  & $22.757\pm0.010$ & $22.413\pm0.010$  \\ 
& WD 28  & $24.048\pm0.010$ & $23.641\pm0.017$ \\ 
& WD 29 & $23.845\pm0.010$ & $23.444\pm0.017 $ \\ 
& WD 30  & $24.156\pm0.010$ & $23.979\pm0.022$  \\ 
& WD 31  & $22.298\pm0.010$ & $22.022\pm0.010$  \\ 
& WD 32  & $24.234\pm0.010$ & $23.838\pm0.020$  \\ 
& WD 33  & $21.803\pm0.010$ & $21.573\pm0.010$  \\ 
& WD 34  & $22.517\pm0.010$ & $22.172\pm0.010$  \\ 
& WD 35  & $23.849\pm0.010$ & $23.299\pm0.019$  \\ \hline
\multicolumn{4}{l}{References: \citet{bellini2010absolute,bellini2010end}}
\end{tabular}
\end{table*}

\begin{table*}
 \caption{Photometry for the WDs in NGC188, NGC2168 and NGC2477.}
\label{tab:phot2}
 \begin{tabular}{ll llll}
 \hline\hline
Cluster & WD & $U\pm\sigma_{U}$ & $B\pm\sigma_{B}$ & $V\pm\sigma_{V}$&$I\pm\sigma_{I}$ \\ \hline
NGC188& WD 1 & $\cdots$ & $\cdots$ & $22.651\pm0.054$ & $22.496\pm0.124$ \\ 
& WD 2 & $\cdots$ & $\cdots$ & $23.350\pm0.119$ & $23.538\pm0.386$ \\ 
& WD 3 & $\cdots$ & $\cdots$ & $23.490\pm0.060$ & $23.413\pm0.185$ \\ 
& WD 4 & $\cdots$ & $\cdots$ & $23.508\pm0.105$ & $23.416\pm0.559$ \\ 
& WD 5 & $\cdots$ & $\cdots$ & $23.669\pm0.045$ & $23.176\pm0.149$ \\ 
& WD 6 & $\cdots$ & $\cdots$ & $24.206\pm0.131$ & $23.756\pm0.228$ \\ 
& WD 7 & $\cdots$ & $\cdots$ & $24.439\pm0.074$ & $24.078\pm0.255$ \\ 
& WD 8 & $\cdots$ & $\cdots$ & $24.469\pm0.054$ & $23.720\pm0.195$ \\ 
& WD 9 & $\cdots$ & $\cdots$ & $25.261\pm0.120$ & $24.422\pm0.297$ \\ \hline
NGC2168& WD 1 & $19.858\pm0.019$ & $20.953\pm0.020$ & $20.989\pm0.019$ & $\cdots$ \\ 
& WD 2 & $20.532\pm0.028$ & $21.63\pm0.031$ & $21.569\pm0.032$ & $\cdots$ \\ 
& WD 3 & $20.506\pm0.020$ & $21.364\pm0.022$ & $21.216\pm0.020$ & $\cdots$ \\ 
& WD 4 & $18.764\pm0.017$ & $19.937\pm0.018$ & $20.065\pm0.017$ & $\cdots$ \\ 
& WD 5 & $18.568\pm0.017$ & $19.735\pm0.017$ & $19.863\pm0.016$ & $\cdots$ \\ 
& WD 6 & $20.339\pm0.025$ & $21.348\pm0.028$ & $21.303\pm0.026$ & $\cdots$ \\ 
& WD 7 & $20.826\pm0.023$ & $21.759\pm0.028$ & $21.701\pm0.027$ & $\cdots$ \\ 
& WD 8 & $19.727\pm0.024$ & $20.746\pm0.024$ & $20.785\pm0.022$ & $\cdots$ \\ 
& WD 9 & $18.720\pm0.018$ & $19.665\pm0.017$ & $19.657\pm0.016$ & $\cdots$ \\ 
& WD 10 & $20.479\pm0.026$ & $21.488\pm0.029$ & $21.398\pm0.026$ & $\cdots$ \\ 
& WD 11 & $20.280\pm0.025$ & $21.650\pm0.031$ & $21.631\pm0.029$ & $\cdots$ \\ 
& WD 12 & $19.558\pm0.019$ & $20.645\pm0.019$ & $20.719\pm0.019$ & $\cdots$ \\ 
& WD 13 & $20.243\pm0.020$ & $21.170\pm0.022$ & $21.175\pm0.020$ & $\cdots$ \\ \hline
NGC2477& WD 1 & $\cdots$ & $\cdots$ & $23.108\pm0.008$ & $22.816\pm0.018$ \\ 
& WD 2 & $\cdots$ & $\cdots$ & $23.689\pm0.010$ & $23.237\pm0.025$ \\ 
& WD 3 & $\cdots$ & $\cdots$ & $23.566\pm0.011$ & $23.282\pm0.024$ \\ 
& WD 4 & $\cdots$ & $\cdots$ & $23.972\pm0.016$ & $23.638\pm0.034$ \\ 
& WD 5 & $\cdots$ & $\cdots$ & $23.957\pm0.015$ & $23.586\pm0.031$ \\ 
& WD 6 & $\cdots$ & $\cdots$ & $23.963\pm0.015$ & $23.576\pm0.030$ \\ 
& WD 7 & $\cdots$ & $\cdots$ & $23.904\pm0.012$ & $23.461\pm0.031$ \\ \hline
\multicolumn{6}{l}{References for NGC188: \citet{von1998wiyn,meibom2009age}}\\
\multicolumn{6}{l}{References for NGC2168: \citet{sung1999ubvi,williams2004}}\\
\multicolumn{6}{l}{References for NGC2477: \citet{jeffery2011the,stein2013}}
\end{tabular}
\end{table*}

% Don't change these lines
\bsp	% typesetting comment

\label{lastpage}
\end{document}